\documentclass[twocolumn,journal]{IEEEtran}
\usepackage[T1]{fontenc}
\usepackage[latin9]{inputenc}
\usepackage{array}
\usepackage{textcomp}
\usepackage{amsmath}
\usepackage{amssymb}
\usepackage{graphicx}
\usepackage[unicode=true,
 bookmarks=true,bookmarksnumbered=true,bookmarksopen=true,bookmarksopenlevel=1,
 breaklinks=false,pdfborder={0 0 0},pdfborderstyle={},backref=false,colorlinks=false]
 {hyperref}
\hypersetup{pdftitle={Your Title},
 pdfauthor={Your Name},
 pdfpagelayout=OneColumn, pdfnewwindow=true, pdfstartview=XYZ, plainpages=false}

\makeatletter

\providecommand{\tabularnewline}{\\}

\let\oldforeign@language\foreign@language
\DeclareRobustCommand{\foreign@language}[1]{%
  \lowercase{\oldforeign@language{#1}}}
\newenvironment{lyxlist}[1]
	{\begin{list}{}
		{\settowidth{\labelwidth}{#1}
		 \setlength{\leftmargin}{\labelwidth}
		 \addtolength{\leftmargin}{\labelsep}
		 }}
	{\end{list}}

\usepackage[caption=false,font=footnotesize]{subfig}

\makeatother

\begin{document}
\title{A Shared-Aperture Dual-Band sub-6 GHz and mmWave Reconfigurable Intelligent
Surface With Independent Operation}
\author{Junhui~Rao,~\IEEEmembership{Student Member,~IEEE,} Yujie~Zhang,~\IEEEmembership{Member,~IEEE,}
Shiwen~Tang,~\IEEEmembership{Student Member,~IEEE,} Zan~Li,~\IEEEmembership{Student Member,~IEEE,}
Zhaoyang Ming, \IEEEmembership{Student Member, IEEE,} Jichen Zhang,
\IEEEmembership{Student Member, IEEE,} Chi Yuk~Chiu,~\IEEEmembership{Senior Member,~IEEE,}
and ~Ross~Murch,~\IEEEmembership{Fellow,~IEEE}\thanks{This work was supported by Hong Kong Research Grants Council Collaborative
Research Fund C6012-20G.}\thanks{Junhui Rao, Zan Li, Zhaoyang Ming, Jichen Zhang, and Chi Yuk Chiu
are with the Department of Electronic and Computer Engineering, the
Hong Kong University of Science and Technology, Hong Kong. (e-mail:
\protect\href{mailto:jraoaa@connect.ust.hk}{jraoaa@connect.ust.hk}).}\thanks{Yujie Zhang and Shiwen~Tang were with the Department of Electronic
and Computer Engineering, The Hong Kong University of Science and
Technology, Hong Kong, and now with the Department of Electrical and
Computer Engineering, National University of Singapore, Singapore.}\thanks{R. Murch is with the Department of Electronic and Computer Engineering
and Institute for Advanced Study (IAS) at the Hong Kong University
of Science and Technology, Hong Kong.     (e-mail: \protect\href{http://eermurch@ust.hk}{eermurch@ust.hk}).}}
\markboth{}{Junhui Rao \MakeLowercase{\emph{et al.}}: babba}
\maketitle
\begin{abstract}
A novel dual-band reconfigurable intelligent surface (DBI-RIS) design
that combines the functionalities of millimeter-wave (mmWave) and
sub-6 GHz bands within a single aperture is proposed. This design
aims to bridge the gap between current single-band reconfigurable
intelligent surfaces (RISs) and wireless systems utilizing sub-6 GHz
and mmWave bands that require RIS with independently reconfigurable
dual-band operation. The mmWave element is realized by a double-layer
patch antenna loaded with 1-bit phase shifters, providing two reconfigurable
states. An 8$\times$8 mmWave element array is selectively interconnected
using three RF switches to form a reconfigurable sub-6 GHz element
at 3.5 GHz. A suspended electromagnetic band gap (EBG) structure is
proposed to suppress surface waves and ensure sufficient geometric
space for the phase shifter and control networks in the mmWave element.
A low-cost planar spiral inductor (PSI) is carefully optimized to
connect mmWave elements, enabling the sub-6 GHz function without affecting
mmWave operation. Finally, prototypes of the DBI-RIS are fabricated,
and experimental verification is conducted using two separate measurement
testbeds. The fabricated sub-6 GHz RIS successfully achieves beam
steering within the range of $-35^{\circ}$ to $35^{\circ}$ for DBI-RIS
with 4$\times$4 sub-6 GHz elements, while the mmWave RIS demonstrates
beam steering between $-30^{\circ}$ to $30^{\circ}$ for DBI-RIS
with 8$\times$8 mmWave elements, and have good agreement with simulation
results.
\end{abstract}


\section{Introduction}

\IEEEPARstart{R}{econfigurable} intelligent surfaces (RISs) have
emerged as a promising technology for future 6th generation (6G) communication
networks \cite{Akyildiz2020,Basharat2022,Liu2020,Yang2019}. They
have garnered significant attention in recent years due to their ability
to intelligently control the electromagnetic propagation environment
\cite{Kaina2014,DiRenzo2020,Huang2019,Subrt2012,Hu2018}. RISs are
planar structures consisting of passive elements that can temporally
modify their electromagnetic properties. This unique characteristic
allows control over the propagation of radio waves in their vicinity,
enabling functions such as beamforming and interference nulling. The
application of RISs has demonstrated improvements in wireless system
performance and also sensing capabilities. Moreover, RISs can be seamlessly
integrated with various emerging communication technologies, including
non-orthogonal multiple access (NOMA), mobile edge computing, unmanned
aerial vehicle (UAV) communication, vehicular networks and physical
layer security \cite{Basharat2022}. Such integration serves to enhance
performance and introduce novel functionalities to these evolving
communication systems.

\begin{figure}[t]
\begin{centering}
\textsf{\includegraphics[width=1\columnwidth]{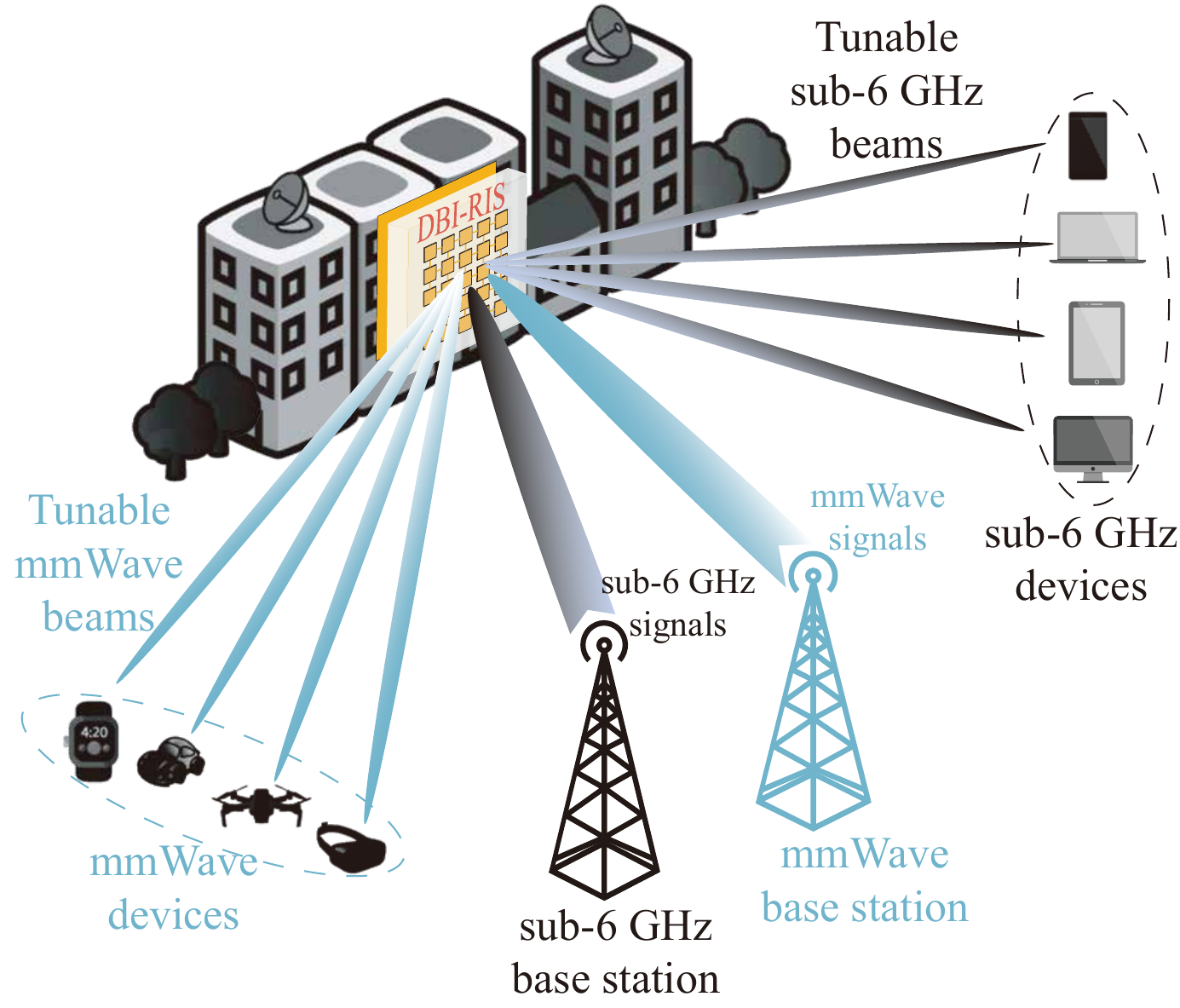}}
\par\end{centering}
\caption{Illustration of the application of the proposed DBI-RIS at sub-6 GHz
and mmWave frequencies.}
\label{DBI-RIS application}
\end{figure}

Various RIS designs for wireless communication have been proposed
in recent years \cite{Pan2021,Rao2023,Rao2022,Zhang2020,Callaghan2021}.
However, most of them primarily focus on a single frequency band,
such as sub-6 GHz or millimeter-wave (mmWave). With the deployment
of 5th generation (5G) communication networks and the development
of 6G systems, more frequency bands, including sub-6 GHz bands and
mmWave bands, have been gradually introduced to fulfill the increasing
requirements of communication. Furthermore, future communication systems
are expected to employ multiple frequency bands simultaneously for
ubiquitous connectivity and massive capacity \cite{Letaief2019,Huang2017}.
Therefore, RIS designs that can simultaneously and independently operate
at both sub-6 GHz and mmWave bands can \textquotedbl double\textquotedbl{}
the capability of RIS. In this paper, we refer to this type of RIS
as dual-band independent RIS (DBI-RIS). As shown in Fig. \ref{DBI-RIS application},
by introducing DBI-RIS in the environment, it can simultaneously and
independently configure the sub-6 GHz and mmWave signals, such as
forming separate beams towards intended devices for each band.

A related technology, dual-band antennas, can be leveraged as a design
methodology for DBI-RIS \cite{Liu2012,Mobashsher2010,Quan2012,Ren2008}.
There are mainly two methods to achieve dual-band antennas: separated-aperture
and shared-aperture. The separated-aperture method forms two independent
antennas in adjacent locations \cite{Zhihong2013,Wang2016}, while
the shared-aperture integrates two antennas with dual-band functionality
into the same aperture, leading to higher aperture efficiency \cite{Hong2017,Xiang2018}.
Additionally, shared-aperture antennas can be further divided into
two main categories: the stacked topology and structure reusing. The
former realizes shared aperture by stacking antenna structures of
different bands in the same aperture, resulting in a higher profile.
For example, in \cite{Zhou2013}, a mmWave patch array was stacked
above an sub-6 GHz patch antenna. Structure reusing aims to utilize
the same structure for both bands twice, thus leading to lower antenna
profile, even though the design can be challenging. Therefore, to
achieve high aperture efficiency and low structure height, shared-aperture
and planar structure reusing is preferred for the design of DBI-RIS.
However, due to the high requirements of simultaneous and independent
reconfigurability in both bands, DBI-RIS designs pose greater challenges.

The first challenge in DBI-RIS design is to ensure independence between
elements for both frequency bands. To fully exploit the capability
of the DBI-RIS in wireless networks, it is crucial that the elements
for the two bands operate independently. This means that the design
must be carefully optimized to avoid unintended effects on the properties
of one band when tuning the properties of the other band. Another
challenge arises from developing elements for two bands in a shared-aperture
manner to maximize the aperture efficiency of the DBI-RIS; integrating
two reconfigurable structures in a single aperture presents difficulties.
The third challenge is balancing the reconfigurability with the complexity
of the control network. RIS technology requires a high degree of reconfigurability
with multiple tunable states to attain optimal performance. However,
this high reconfigurability requires a complex control network, particularly
for the shared-aperture DBI-RIS, where two independent control networks
are needed within a limited area. There is a trade-off between the
degree of reconfigurability and the complexity of the control network,
which needs to be carefully balanced. The final challenge involves
prototyping and conducting experimental verification. The DBI-RIS
features a complex structure for the mmWave frequency band, which
is sensitive to fabrication errors. Therefore, potential inevitable
errors must be included in the design process. Additionally, to validate
the performance of the DBI-RIS, two independent measurement setups
are required at each frequency band for experimental verification.

To meet the requirements for RIS operation in both sub-6 GHz and mmWave
bands, a shared-aperture DBI-RIS is proposed in this paper with a
comprehensive investigation into its design methodology, fabrication,
and measurement. The contributions and novelties of this paper can
be summarized as follows:

\textit{1) Architecture of DBI-RIS:} A DBI-RIS based on structure-reusing
shared-aperture configuration is proposed and targeted for sub-6 GHz
(3.5 GHz) and mmWave (28 GHz) frequency bands. The mmWave function
is realized by an array of double-layer patch antennas loaded by 1\textendash bit
phase shifters, while the sub-6 GHz function is achieved by the selective
interconnection of the mmWave patches with 3-bit reconfigurability.
The entire aperture is utilized for both sub-6 GHz and mmWave without
wasting space. The independence between two frequency bands is ensured,
and two independent controlling networks are also designed in the
compact area.

\textit{2) Suspended electromagnetic band gap (EBG) structure:} Due
to the integration of mmWave and sub-6 GHz functions, the thickness
of the substrate is much thicker than a normal mmWave patch antenna,
leading to strong surface waves and causing irregular radiation patterns
for the mmWave antenna. Therefore, suspended EBG is proposed in this
paper to not only suppress the surface waves but also ensure enough
geometric space for the phase shifter and controlling networks.

\textit{3) Planar spiral inductor (PSI):} To reuse the mmWave structures
for the sub-6 GHz function, the mmWave patches have to be selectively
connected together to form a specific geometric pattern. However,
direct connection by metal wires will inevitably destroy the mmWave
function. Therefore, PSI is proposed and carefully optimized in this
paper to enable the sub-6 GHz function without affecting the operation
of mmWave.

\textit{4) Prototype and Experimental Verification: }A prototype of
the proposed DBI-RIS is fabricated, and two separate measurement testbeds
were used to test the scattered patterns and verify effectiveness
of the DBI-RIS in each band.

This paper is organized as follows. Section II presents the architecture
of the DBI-RIS, providing an overview of the proposed design. Section
III describes the working methodology of the mmWave RIS, including
the design of the single mmWave element, mmWave array, suspended EBG
and supporting experimental results. Section IV outlines the working
methodology of the sub-6 GHz RIS, covering the design of the single
sub-6 GHz element, PSI and supporting experimental results. In Section
V, we discuss the integration of sub-6 GHz elements into a 4$\times$4
RIS configuration, which shares the aperture with a 32$\times32$
element RIS operating in the mmWave band. Section VI compares the
proposed design with related work and provides discussions. Finally,
Section VII concludes this work.

\section{DB-RIS Architecture}

\begin{figure}[t]
\begin{centering}
\textsf{\includegraphics[width=1\columnwidth]{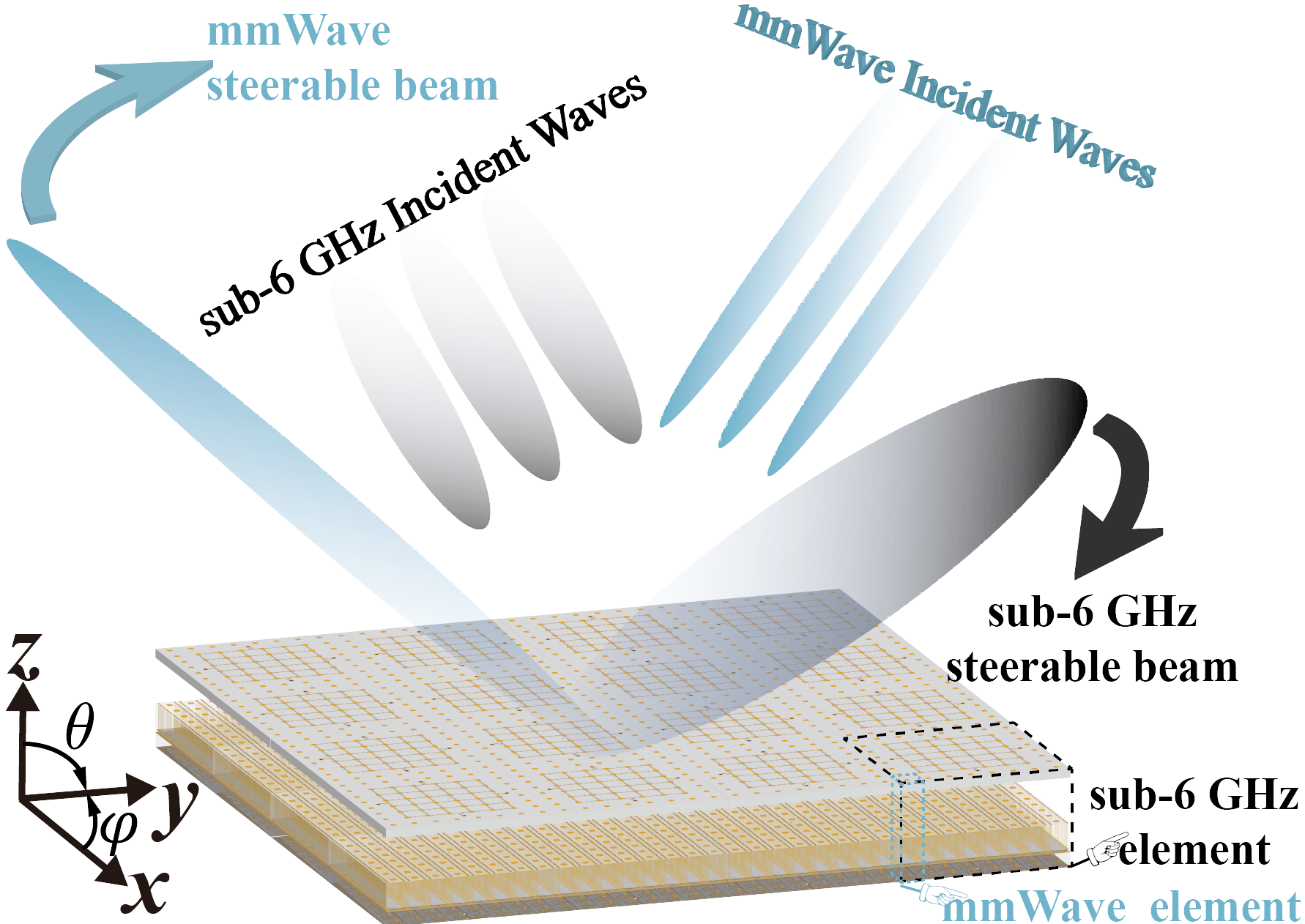}}
\par\end{centering}
\caption{Illustration of the proposed DBI-RIS constructed from 4$\times$4
sub-6 GHz elements and 32$\times$32 mmWave elements. The coordinate
system for the DBI-RIS is also shown and used throughout this paper.
In the coordinate system shown on the DBI-RIS, we refer to azimuth
as being the angle $\varphi$ and elevation being the angle $\theta$.
The detailed geometry of a single sub-6 GHz element is provided in
Fig. \ref{single sub-6GHz element}.}
\label{4x4 sub-6 RIS}
\end{figure}

\begin{figure}[tbh]
\begin{centering}
\textsf{\includegraphics[width=1\columnwidth]{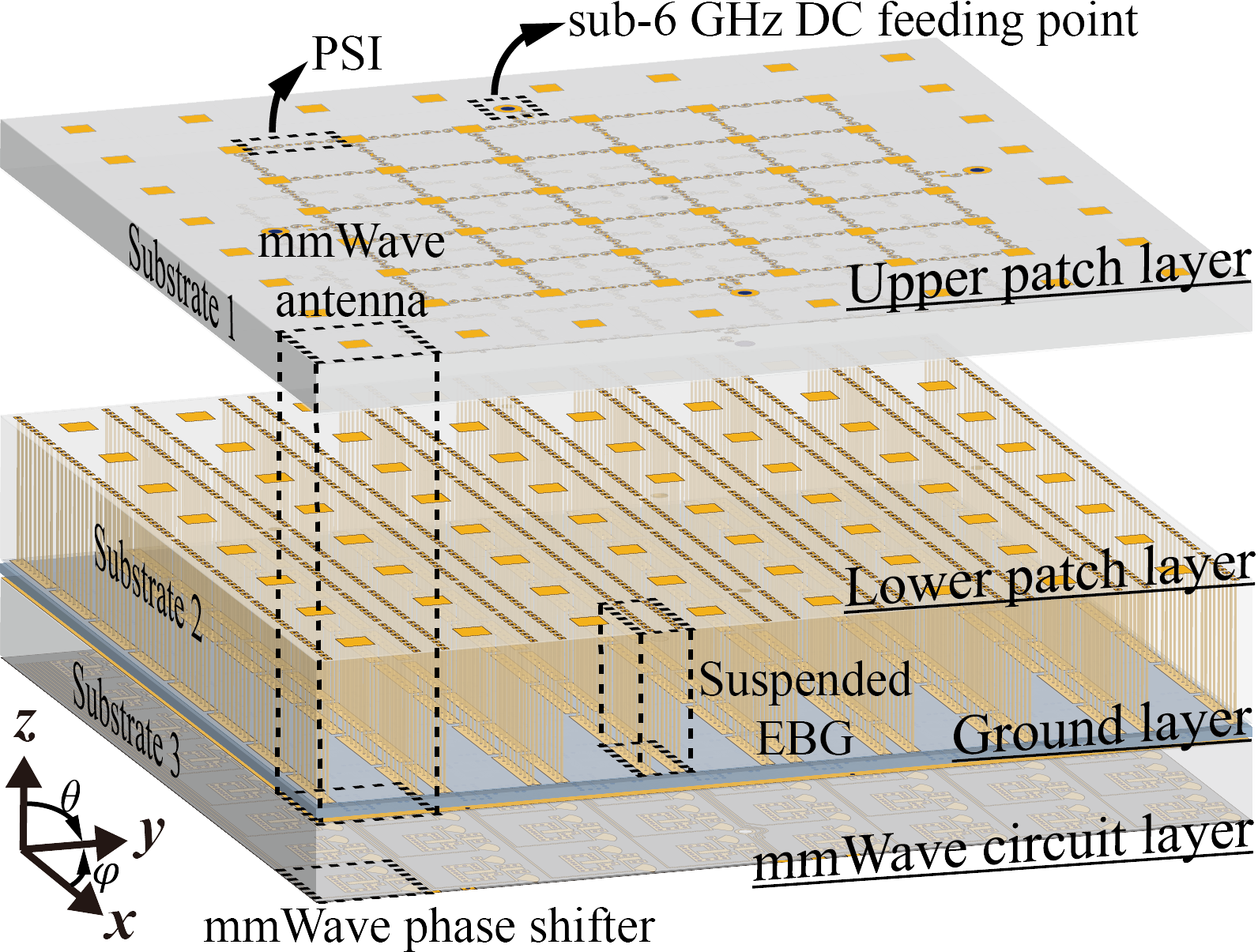}}
\par\end{centering}
\begin{centering}
(a)
\par\end{centering}
\begin{centering}
\textsf{\includegraphics[width=0.5\columnwidth]{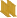}\includegraphics[width=0.5\columnwidth]{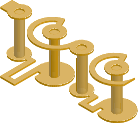}}
\par\end{centering}
\begin{raggedright}
\hspace*{0.24\columnwidth}(b)\hspace*{0.44\columnwidth} (c)
\par\end{raggedright}
\caption{(a) Illustration of a single sub-6 GHz element consisting of 8$\times$8
mmWave elements with detailed structures. (b) Magnified view of the
suspended EBG. (c) Magnified view of the PSI. }
\label{single sub-6GHz element}
\end{figure}

Fig. \ref{4x4 sub-6 RIS} illustrates the overall architecture of
the proposed DBI-RIS while Fig. \ref{single sub-6GHz element} (a)
provides a detailed view of the sub-6 GHz element, as well as the
8$\times$8 mmWave elements inside it. In the design shown in Fig.
\ref{4x4 sub-6 RIS}, the RIS contains 4$\times$4 sub-6 GHz elements,
and each sub-6 GHz element shares its aperture with 8$\times$8 mmWave
elements providing 32$\times$32 mmWave elements in total. It is worth
noting that the DBI-RIS can be constructed using any desired number
of sub-6 GHz elements and corresponding mmWave elements, enabling
flexibility in size and performance. Both the sub-6 GHz and mmWave
elements are carefully designed to meet the requirements of RIS, including
effective performance across a wide angular range, sufficient bandwidth,
and the ability to temporally adjust the phase shifts of incident
waves. Importantly, the proposed sub-6 GHz and mmWave elements maintain
independence from each other. This means that when both mmWave and
sub-6 GHz signals impinge on the DBI-RIS, the system can independently
operate for each frequency band, allowing for separate beamforming
capabilities, as exemplified in Fig. \ref{4x4 sub-6 RIS}. The target
frequency bands for sub-6 GHz and mmWave are 3.5 GHz and 28 GHz, respectively,
but they can also be scaled to other sub-6 GHz and mmWave bands.

\begin{figure}[th]
\begin{centering}
\textsf{\includegraphics[width=0.75\columnwidth]{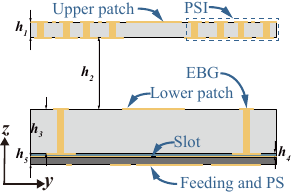}}
\par\end{centering}
\begin{centering}
(a)
\par\end{centering}
\begin{centering}
\textsf{\includegraphics[width=0.4\columnwidth]{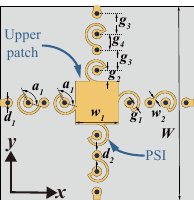}}\hspace*{0.1\columnwidth}\textsf{\includegraphics[width=0.4\columnwidth]{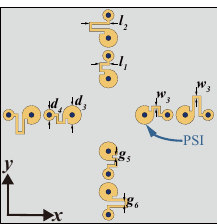}}
\par\end{centering}
\begin{raggedright}
\hspace*{0.23\columnwidth}(b)\hspace*{0.44\columnwidth} (c)
\par\end{raggedright}
\begin{centering}
\textsf{\includegraphics[width=0.4\columnwidth]{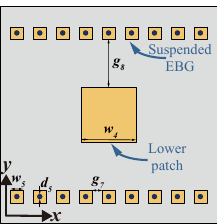}}\hspace*{0.1\columnwidth}\textsf{\includegraphics[width=0.4\columnwidth]{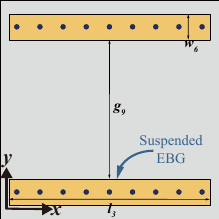}}
\par\end{centering}
\begin{raggedright}
\hspace*{0.23\columnwidth}(d)\hspace*{0.44\columnwidth} (e)
\par\end{raggedright}
\begin{centering}
\textsf{\includegraphics[width=0.4\columnwidth]{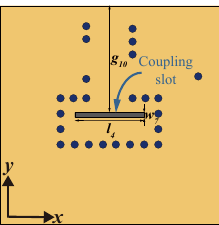}}\hspace*{0.1\columnwidth}\textsf{\includegraphics[width=0.4\columnwidth]{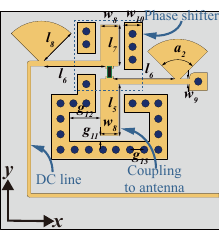}}
\par\end{centering}
\begin{raggedright}
\hspace*{0.23\columnwidth}(f)\hspace*{0.44\columnwidth} (g)
\par\end{raggedright}
\caption{Orthographic views of a single mmWave element. (a) Cross-sectional
view. (b) Upper patch layer. (c) Bottom layer of the substrate 1.
(d) Lower patch layer. (e) Bottom layer of the substrate 2. (f) Ground
layer. (g) mmWave circuit layer. The dimensions are $h_{1}=0.508$,
$h_{2}=2.5$, $h_{3}=1.524$, $h_{4}=0.204$, $h_{5}=0.252$, $W=8.5$,
$a_{1}=205.4$, $a_{2}=90$, $g_{1}=0.12$, $g_{2}=0.473$, $g_{3}=0.897$,
$g_{4}=0.697$, $g_{5}=0.1$, $g_{6}=0.157$, $g_{7}=0.4$, $g_{8}=1.885$,
$g_{9}=5.42$, $g_{10}=4.15$, $g_{11}=0.2$, $g_{12}=1.2$, $g_{13}=0.5$,
$d_{1}=0.2$, $d_{2}=0.45$, $d_{3}=0.74$, $d_{4}=0.45$, $d_{5}=0.2$,
$w_{1}=1.9$, $w_{2}=0.1$, $w_{3}=0.1$, $w_{4}=2.15$, $w_{5}=0.5$,
$w_{6}=1$, $w_{7}=0.2$, $w_{8}=0.7$, $w_{9}=0.2$, $w_{10}=0.7$,
$l_{1}=0.4$, $l_{2}=0.8$, $l_{3}=7.8$, $l_{4}=2.7$, $l_{5}=2$,
$l_{6}=2.4$, $l_{7}=1.6$, $l_{8}=1.51$. (Unit: mm)}
\label{Orthographic views single mm-wave element}
\end{figure}

Referring to Fig. \ref{single sub-6GHz element} (a), it can be seen
that the proposed DBI-RIS consists of three layers of substrates,
as labeled on the left of Fig. \ref{single sub-6GHz element}(a),
where substrates 2 and 3 are laminated together by a bondply, while
substrates 1 and 2 are separated by an air gap. Substrates 1 and 2
are fabricated using Rogers 4003C substrate ($\epsilon_{r}=3.55$
and loss $\mathrm{tangent}=0.0027$), and substrate 3 is fabricated
using Rogers 5880 substrate ($\epsilon_{r}=2.2$ and loss $\mathrm{tangent}=0.0009$).
The detailed dimensions are provided later in Fig. \ref{Orthographic views single mm-wave element}.

The single sub-6 GHz reconfigurable element in Fig. \ref{single sub-6GHz element}
(a) is realized by appropriately connecting 8$\times$8 mmWave elements
together (as detailed later) in the \textquotedbl upper patch layer\textquotedbl .
The reconfiguration of a single mmWave element is achieved by a 1-bit
reflection phase shifter in the \textquotedbl mmWave circuit layer\textquotedbl{}
and a double-layer patch antenna with one patch located at the \textquotedbl lower
patch layer\textquotedbl{} and another located at the \textquotedbl upper
patch layer\textquotedbl . The DC controlling network for mmWave
elements is also located in the ``mmWave circuit layer''. The ground
for both mmWave and sub-6 GHz elements is located at the \textquotedbl ground
layer\textquotedbl , which is the top side of substrate 3. A suspended
EBG is also proposed and located inside substrate 2, with the top
part in the ``lower patch layer'' and it extends into substrate
2. We refer to it as a suspended EBG because it is not directly connected
to the \textquotedbl ground layer\textquotedbl . Fig. \ref{single sub-6GHz element}
(b) demonstrates the magnified view of the proposed suspended EBG.
Reconfigurability of the sub-6 GHz element is formed by selectively
connecting the upper patches of mmWave elements using PSI, which is
a two-layer spiral structure with half in the ``upper patch layer''
and the other half on the bottom of substrate 1 as shown in Fig. \ref{single sub-6GHz element}
(c). Control of the sub-6 GHz element is performed using four DC feeding
points and 3 PIN diode switches in the ``upper patch layer'' and
connect to the mmWave circuit layer using metal pillars for forming
the DC controlling network.

In the following sections, the methodology for the designs of mmWave
element and the sub-6 GHz element are provided. 

\section{mmWave RIS Working Methodology}

\subsection{Architecture of a single mmWave element}

Fig. \ref{Orthographic views single mm-wave element} provides an
orthographic view of a single mmWave element with detailed geometry
and the parameters for each layer. It corresponds to one of the 8
$\times$ 8 mmWave elements inside a sub-6 GHz element (the sub-6
GHz element structure is described later in Section IV). A double-layer
patch antenna and a 1-bit phase shifter are combined together to realize
the mmWave RIS element with two reconfigurable states. Fig. \ref{Orthographic views single mm-wave element}
(a) displays the specific thickness of each substrate, as well as
the air gap. The PSI and suspended EBG structures are also designed
across substrates 1 and 2, respectively. It can also be seen that
the suspended EBG is not connected to the ground. The upper and lower
patches are located in the center of the element, as shown in Fig.
\ref{Orthographic views single mm-wave element} (b) and (d). As shown
in Fig. \ref{Orthographic views single mm-wave element} (b) and (c),
four PSIs are appropriately located on the four sides of the upper
patches for the formation of the sub-6 GHz element, as described later
in Section IV. Two rows of suspended EBG structures are symmetrically
designed in the y-direction with long metallic strips, as shown in
Fig. \ref{Orthographic views single mm-wave element} (e). The feeding
of the mmWave antenna is based on the aperture feeding technique with
a slot in the ground, as shown in Fig. \ref{Orthographic views single mm-wave element}
(f). The 1-bit phase shifter and its controlling DC line are shielded
by the ground and thus do not affect the operation of the mmWave antenna,
as shown in Fig. \ref{Orthographic views single mm-wave element}
(g).

\subsection{mmWave antenna}

The antenna is the receiving and re-radiating component in the mmWave
element. As discussed in \cite{Rao2023}, to realize wide-angle beamforming,
an antenna with a wide beamwidth is preferred. Therefore, the patch
antenna with a relatively wide beam and planar structure is an ideal
choice for RIS. In addition, since the patches will be reused in the
later design of the sub-6 GHz element, we must also consider the performance
of the dual-band elements. mmWave patch antennas require thin substrates
for normal operation to ensure radiation patterns without large side
lobes, while the sub-6 GHz element requires a thick substrate for
good bandwidth. To balance this trade-off, we form the antenna with
a double-layer of patches. Combined with the relatively thick substrate
2 (1.524 mm) and large air gap (2.5 mm), the bandwidth of the sub-6
GHz can be acceptable, and the mmWave antenna can effectively radiate
at 28 GHz with a good radiation pattern in the broadside direction
with the help of the suspended EBG.

\begin{figure}[t]
\begin{centering}
\textsf{\includegraphics[height=0.3\columnwidth]{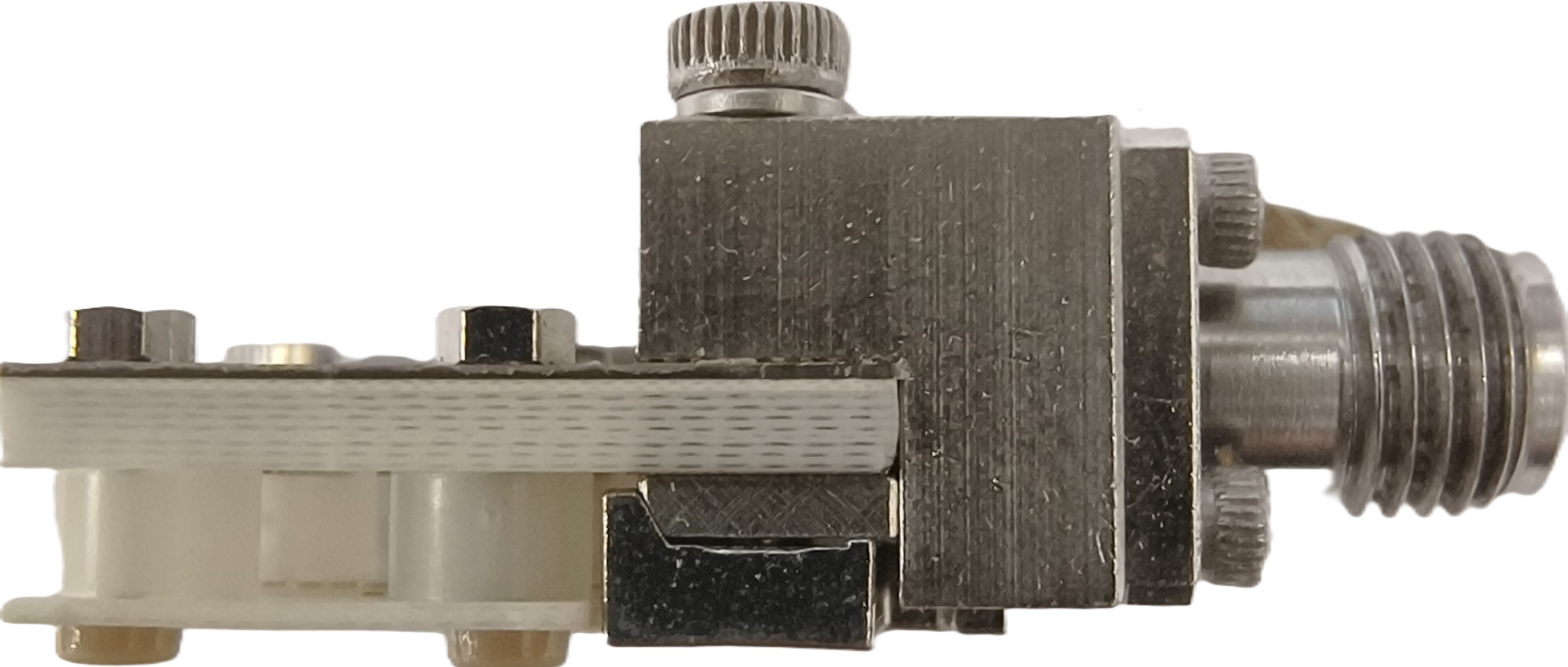}}
\par\end{centering}
\begin{centering}
(a)
\par\end{centering}
\begin{centering}
\textsf{\includegraphics[height=0.25\columnwidth]{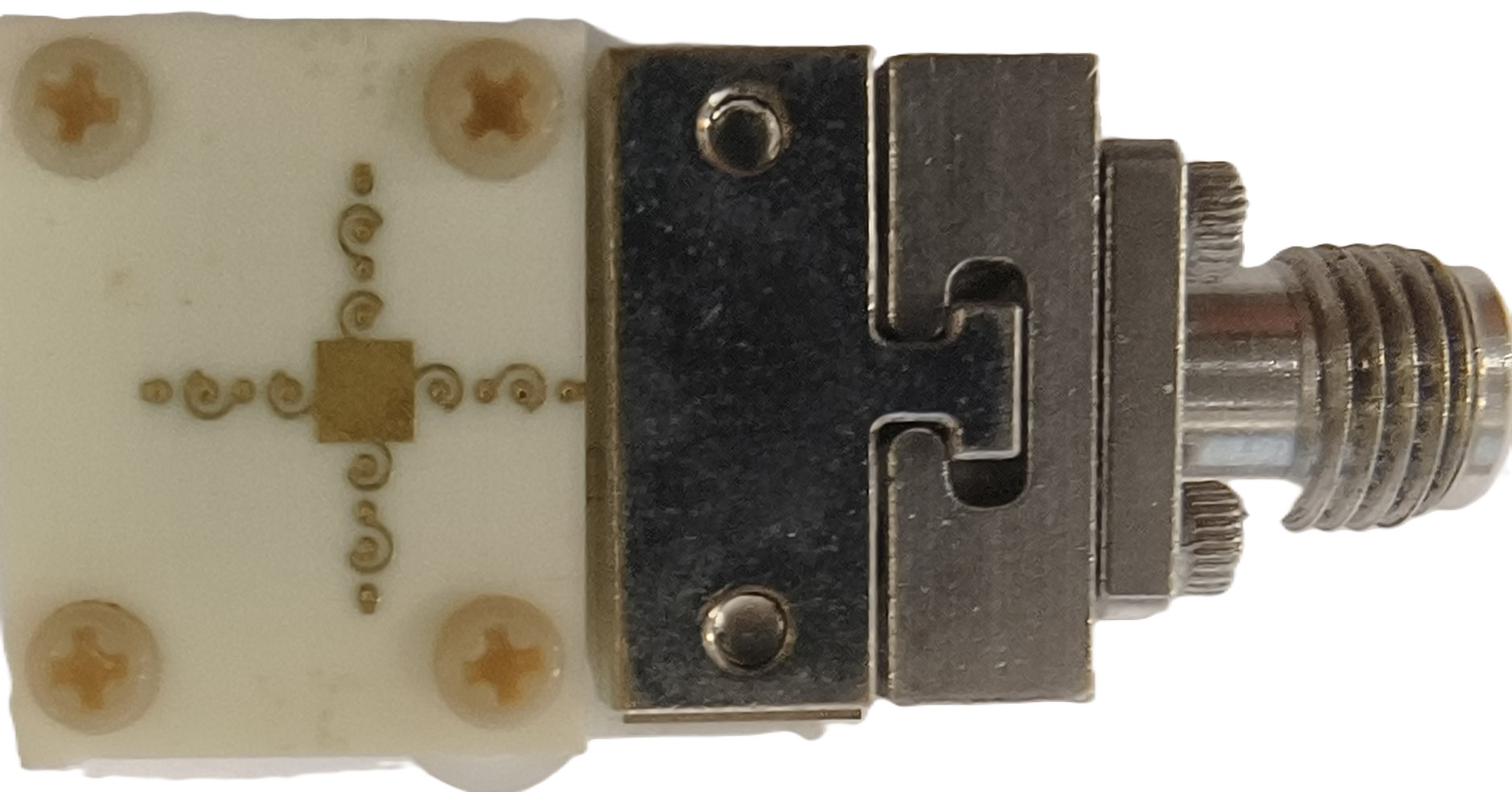}\includegraphics[height=0.25\columnwidth]{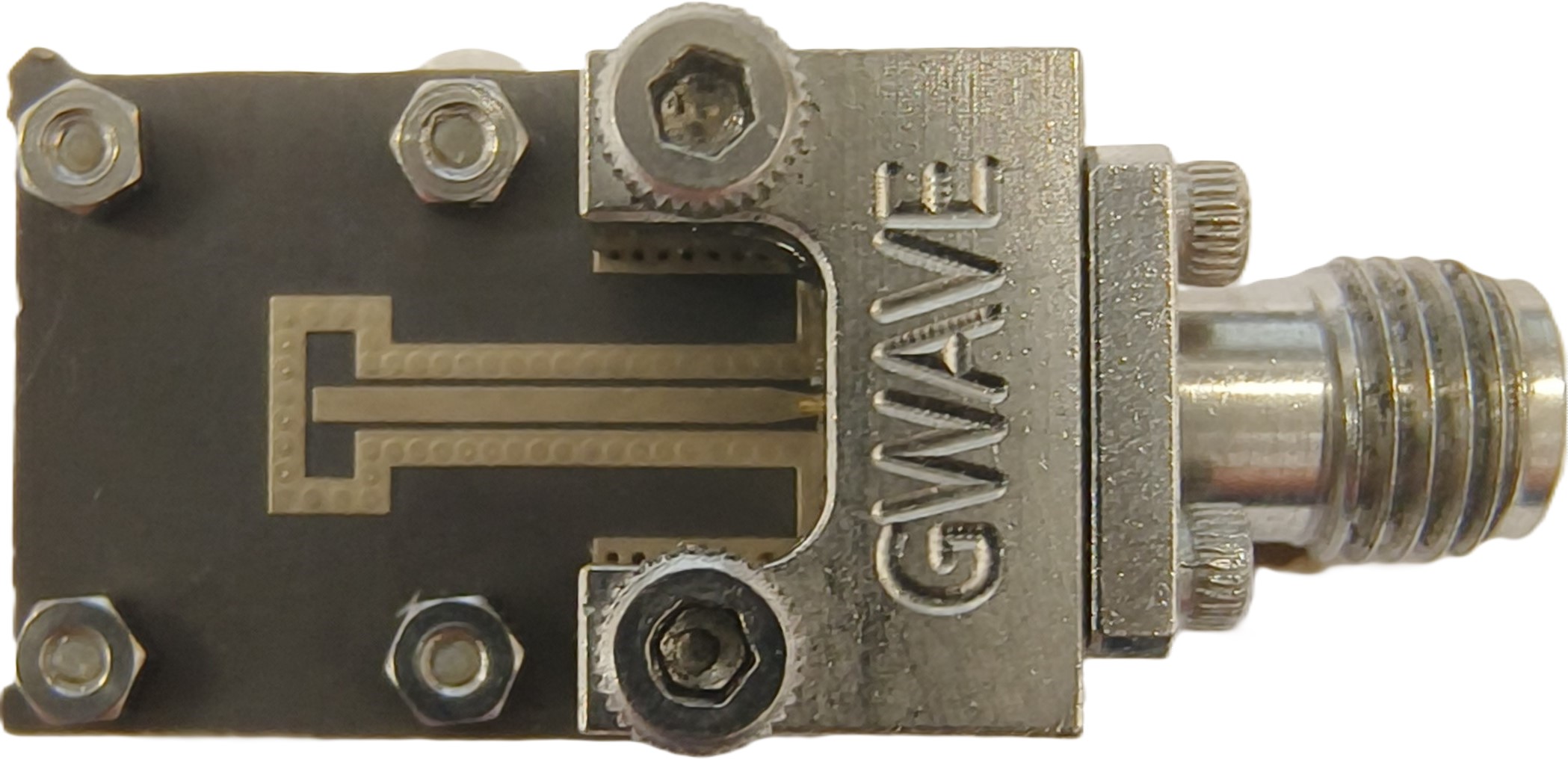}}
\par\end{centering}
\begin{raggedright}
\hspace*{0.23\columnwidth}(b)\hspace*{0.44\columnwidth} (c)
\par\end{raggedright}
\caption{Photograph of the fabricated mmWave antenna with RF connector. (a)
Cross-sectional view. (b) Top view. (c) Bottom view.}
\label{Photograph of the fabricated mm-wave antenna}
\end{figure}

\begin{figure}[t]
\begin{centering}
\textsf{\includegraphics[width=0.5\columnwidth]{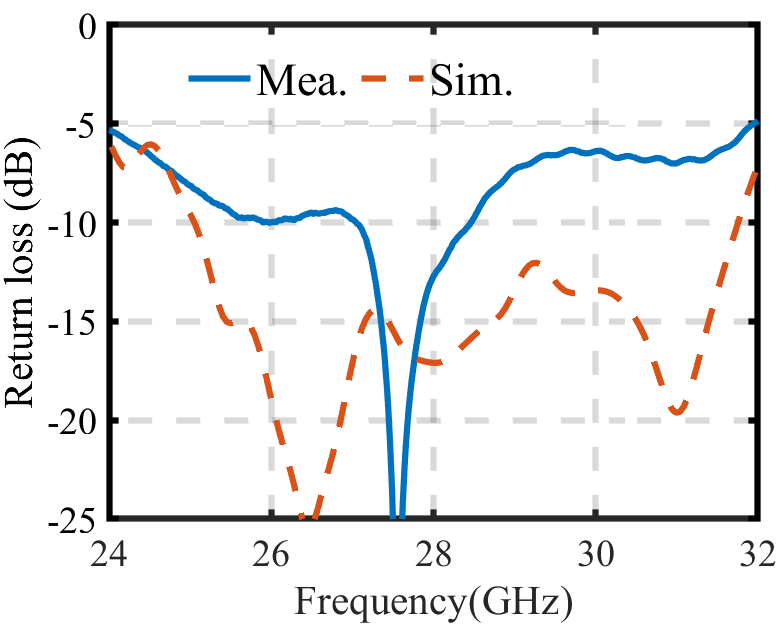}\includegraphics[width=0.5\columnwidth]{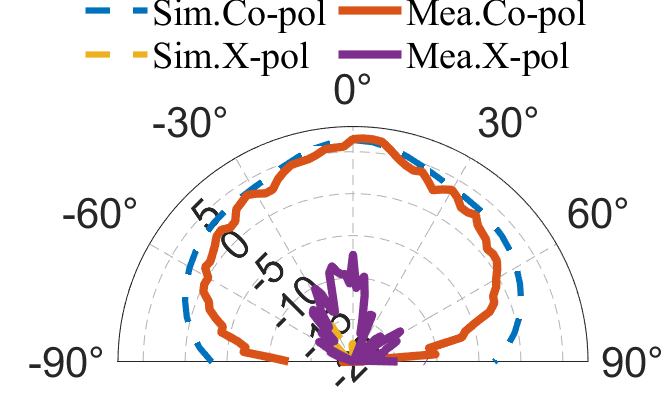}}
\par\end{centering}
\begin{raggedright}
\hspace*{0.23\columnwidth}(a)\hspace*{0.44\columnwidth} (b)
\par\end{raggedright}
\caption{Simulated and measured performance of the fabricated mmWave antenna
with (a) return loss, and (b) radiation patterns at 28 GHz (Unit:
dBi). }
\label{performance of mm-wave antenna}
\end{figure}

A prototype without the phase shifter and with an RF connector, Gwave
SMA-KFD0851, is shown in Fig. \ref{Photograph of the fabricated mm-wave antenna}.
The substrates are tightly fixed with screws, and nylon rings are
utilized to ensure the accurate air gap. A vector network analyzer,
Rohde \& Schwarz ZNA67, was used to measure the return loss of the
fabricated antenna. Fig. \ref{performance of mm-wave antenna} demonstrates
the simulated and measured performance of the mmWave antenna. The
measured operating bandwidth (return loss < -10 dB) ranges from 27.05
GHz to 28.45 GHz, and the measured radiation patterns show the antenna
can effectively radiate in the broadside direction with 6.5 dBi gain.
During the measurement process, the mmWave antenna is placed above
an absorber for better measurement, thus the back lobe of the pattern
was not measured as shown. Besides, at mmWave bands, small fabrication
errors and interconnections with SMA connectors will have a relatively
large impact on the performance of the antenna, resulting in noticeable
discrepancies in the measured results. However, even so, the fabricated
mmWave antenna still operates effectively with good performance for
our application.

\subsection{Phase shifter}

\begin{figure}[t]
\begin{centering}
\textsf{\includegraphics[height=0.4\columnwidth]{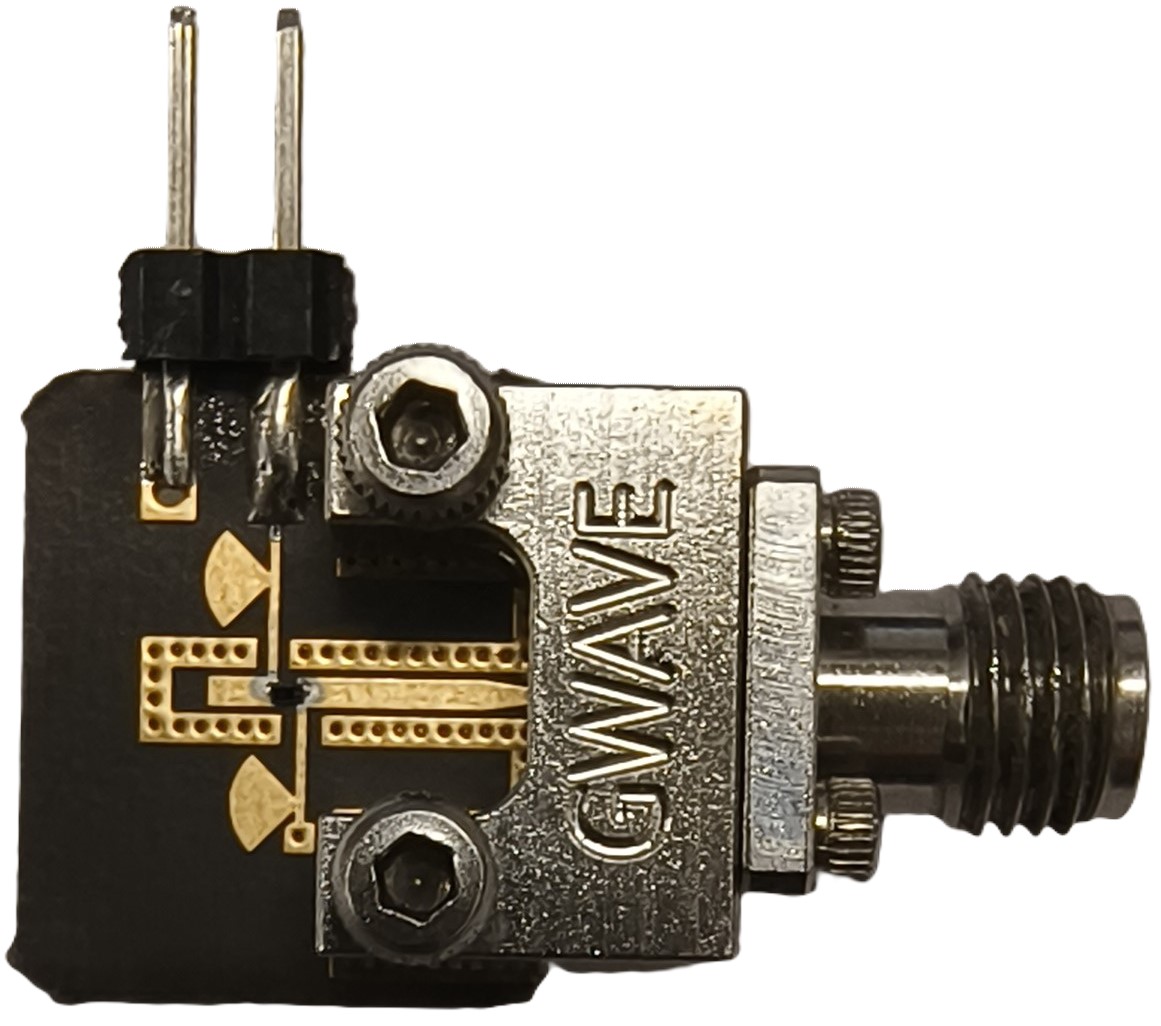}\includegraphics[height=0.4\columnwidth]{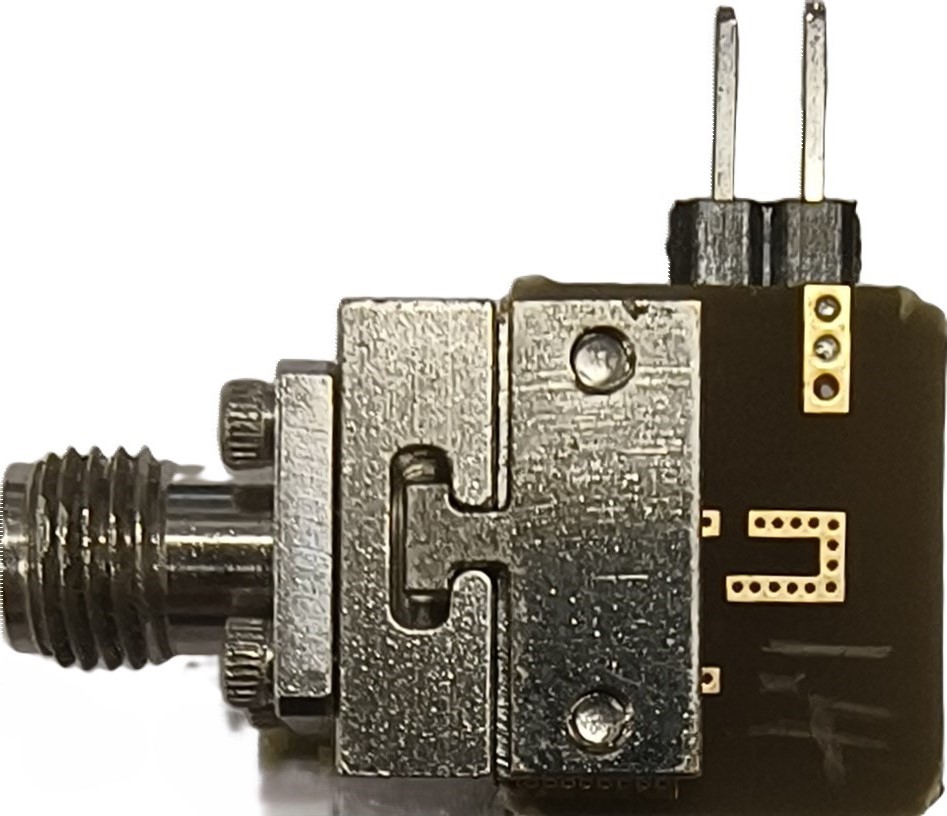}}
\par\end{centering}
\begin{raggedright}
\hspace*{0.23\columnwidth}(a)\hspace*{0.44\columnwidth} (b)
\par\end{raggedright}
\caption{Photograph of the fabricated mmWave phase shifter with RF connector.
(a) Top view. (b) Bottom view.}
\label{Picture mmwave PS}
\end{figure}

To realize the phase reconfigurability function in the mmWave element,
a 1-bit reflection phase shifter is proposed, designed, and fabricated
as shown in Fig. \ref{Orthographic views single mm-wave element}
(g) and Fig. \ref{Picture mmwave PS} without the mmWave antenna.
As a reconfigurable component, the complexity of the phase shifter
depends on the resolution or control bits. Increasing the number of
bits necessitates a higher number of adjustable states, additional
control lines, and a more intricate control system for the RIS. This
becomes especially important when scaling the size of an RIS, as the
routing of control lines and the complexity of the controlling system
significantly increases with the inclusion of more elements. In addition,
for the mmWave band, a large number of elements are expected to be
integrated in a compact area due to the small physical aperture of
each element. Thus, 1-bit is the feasible choice for the phase shifter
as a balance between complexity and reconfigurability.

\begin{figure}[t]
\begin{centering}
\textsf{\includegraphics[width=0.5\columnwidth]{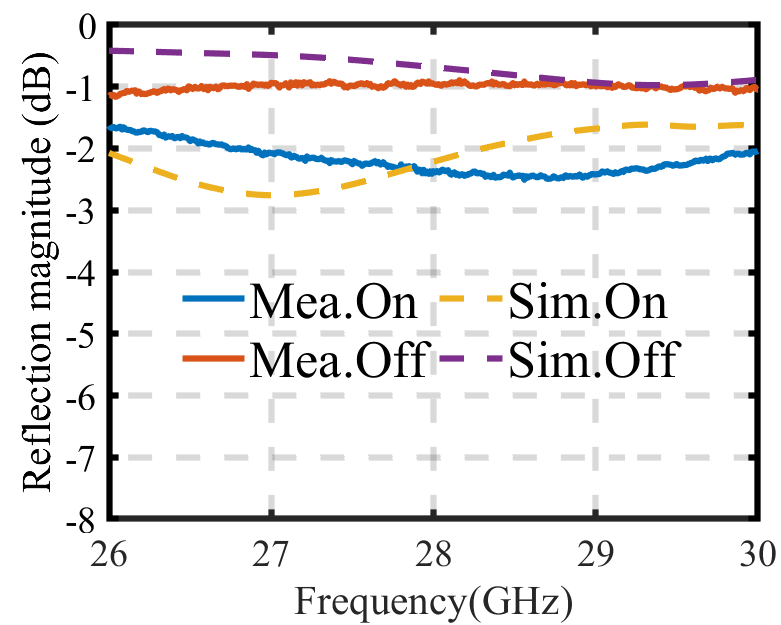}\includegraphics[width=0.5\columnwidth]{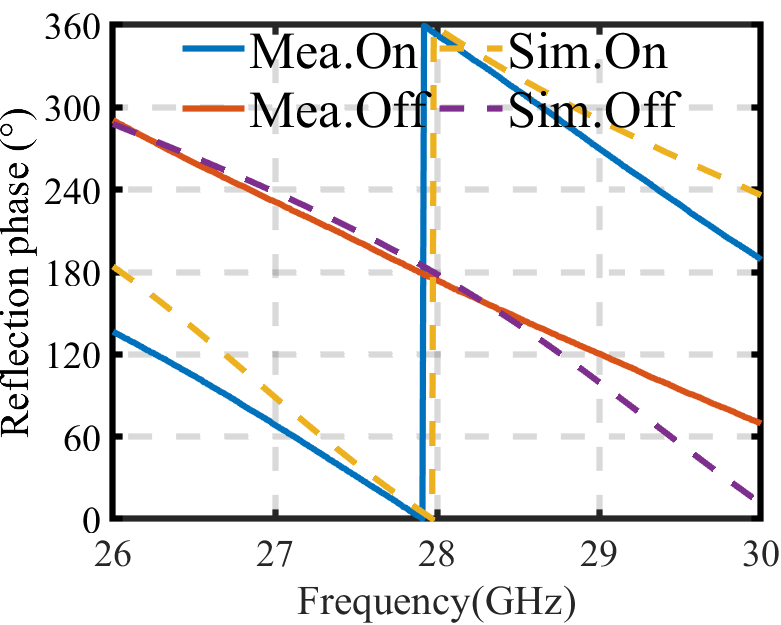}}
\par\end{centering}
\begin{raggedright}
\hspace*{0.23\columnwidth}(a)\hspace*{0.44\columnwidth} (b)
\par\end{raggedright}
\caption{Simulated and measured performance of the fabricated mmWave phase
shifter with (a) reflection magnitude, and (b) reflection phase. }
\label{performance mmwave PS}
\end{figure}

The proposed phase shifter is designed by inserting a diode, MACOM
MA4GP907, in the middle of an open-ended transmission line. By adjusting
the length of the separated line, the reflected phases when the diode
is on and off can be tuned to the desired values. Two sector chokes
are also designed on the two sides of the diode for DC feeding. Fig.
\ref{performance mmwave PS} presents the measured and simulated results
of the prototyped phase shifter in terms of magnitude and phases.
The measured reflection magnitude is about -2.2 dB and -1 dB for the
off and on states, respectively, and the phase difference between
the two states is successfully adjusted to $180^{\circ}$. The small
discrepancies between measured and simulated results are potentially
caused by fabrication errors and the effect of the SMA connector.

\subsection{Suspended EBG}

\begin{figure}[t]
\begin{centering}
\textsf{\includegraphics[height=0.33\columnwidth]{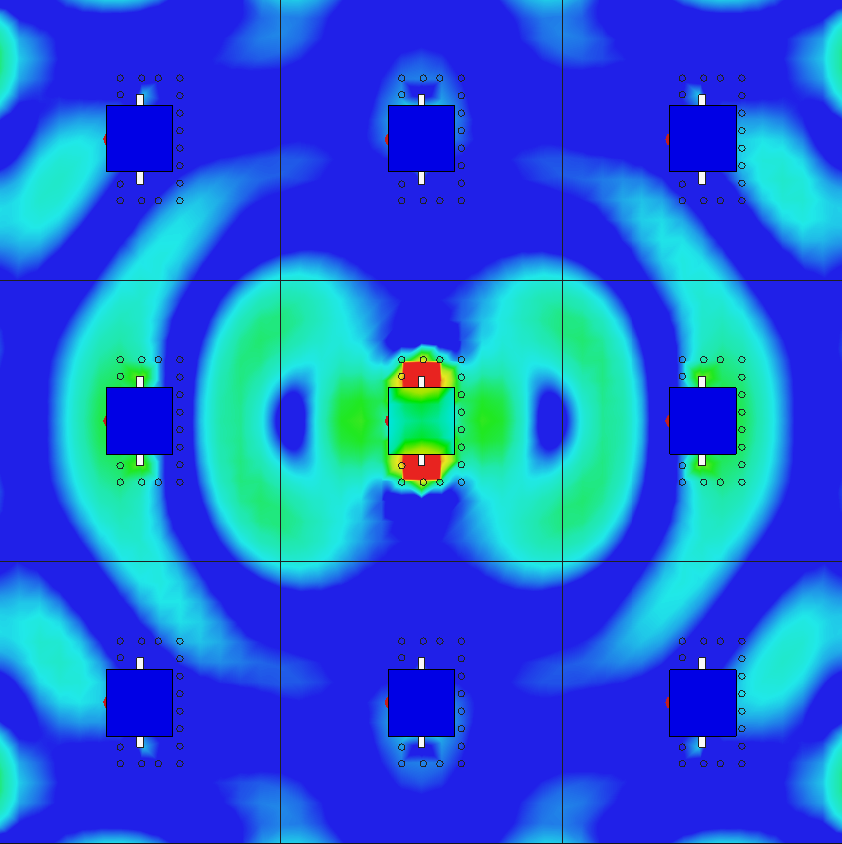}\includegraphics[height=0.33\columnwidth]{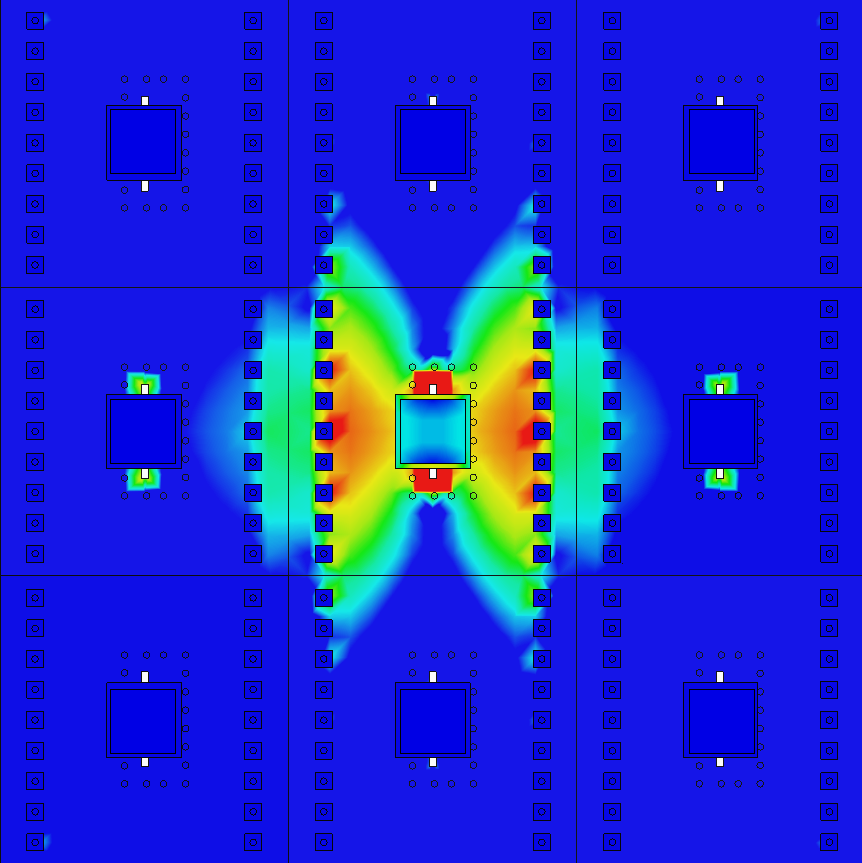}\includegraphics[height=0.33\columnwidth]{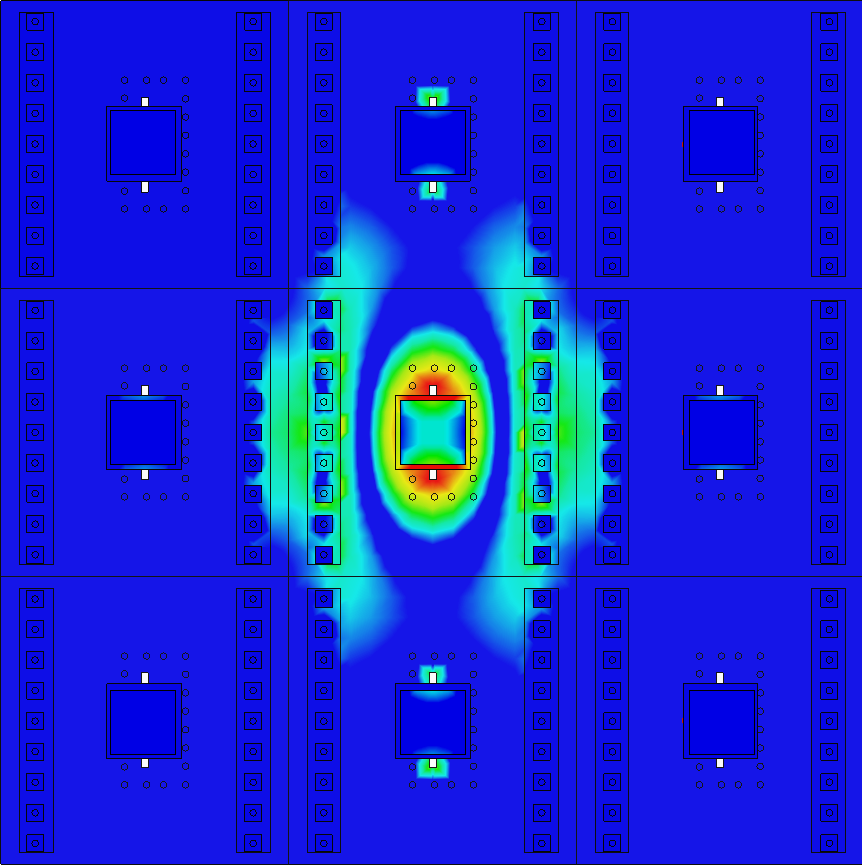}}
\par\end{centering}
\begin{raggedright}
\hspace*{0.15\columnwidth}(a)\hspace*{0.27\columnwidth} (b)\hspace*{0.27\columnwidth}
(c)
\par\end{raggedright}
\caption{The simulated surface current distribution of 3$\times$3 mmWave antenna
array when the central antenna is excited with (a) no EBG, (b) traditional
EBG, and (c) suspended EBG.}
\label{surface current distribution}
\end{figure}

\begin{figure}[t]
\begin{centering}
\textsf{\includegraphics[width=0.5\columnwidth]{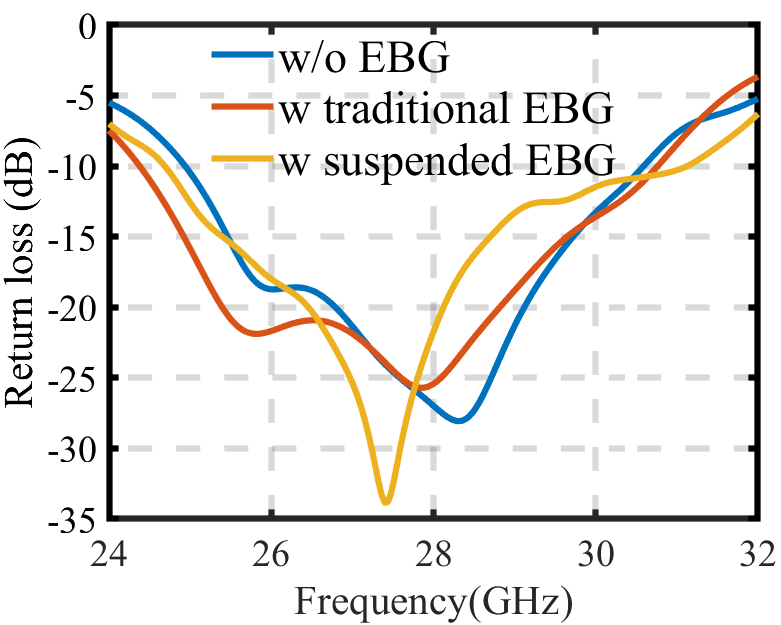}\includegraphics[width=0.5\columnwidth]{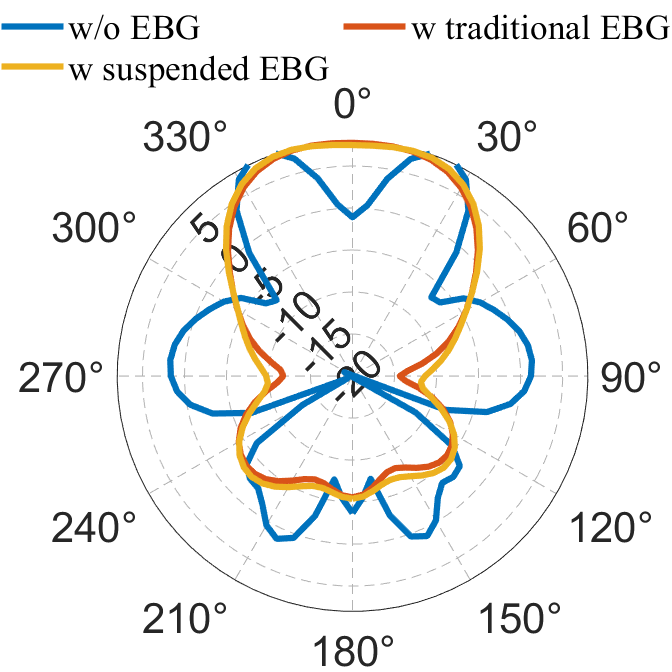}}
\par\end{centering}
\begin{raggedright}
\hspace*{0.23\columnwidth}(a)\hspace*{0.44\columnwidth} (b)
\par\end{raggedright}
\caption{Simulated performance of the central antenna in a 3$\times$3 mmWave
antenna array for different EBG structures with (a) return loss, and
(b) radiation patterns at 28 GHz (Unit: dBi). The PSI is not included
in all simulations here for simplicity.}
\label{performance of different EBG}
\end{figure}

As discussed in Section III.B, to ensure sufficient bandwidth, the
thickness of substrate 2 and the air gap are designed to be large
in terms of the mmWave band. This leads to a strong surface wave inside
the substrate. Thus, even though the antenna can still radiate, the
radiation pattern becomes irregular and shows significant side lobes
in the end-fire directions, as shown in Fig. \ref{surface current distribution}
(a) and Fig. \ref{performance of different EBG} (b). However, a radiation
pattern with low side-lobe levels in the undesired directions and
wide beamwidth in the broadside direction is important for the performance
of the RIS \cite{Rao2023}. 

EBG can thus be utilized to suppress the surface wave and ensure the
low side-love level  of the radiation pattern \cite{Llombart2005}.
As demonstrated in Fig. \ref{surface current distribution} (b), two
rows of traditional EBG are designed on both sides of the patch and
in the orthogonal direction to the propagation direction of the surface
wave. By adjusting the key geometric parameters, including the gap
between adjacent mushrooms and the distance between the patch and
traditional EBG, the surface wave can be effectively suppressed, and
the side lobes in the end-fire directions of radiation pattern is
also suppressed with good return loss performance, as shown in Fig.
\ref{performance of different EBG}. 

\begin{figure}[t]
\begin{centering}
\textsf{\includegraphics[width=1\columnwidth]{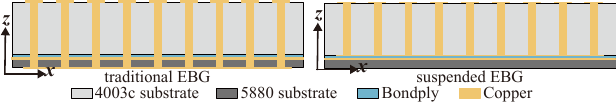}}
\par\end{centering}
\caption{Cross-sectional view of the traditional EBG and the proposed suspended
EBG.}
\label{cross view EBG}
\end{figure}

However, in order to directly connect the vias of traditional EBG
to the ground layer, vias have to penetrate through the ground and
reach the mmWave circuit layer due to the limitations of the multi-layer
PCB fabrication process, as demonstrated in Fig. \ref{cross view EBG}.
On the other hand, the space in the mmWave circuit layer is extremely
limited. The phase shifter and controlling network occupy most of
the area, and there is not enough space for the penetrating vias and
corresponding pads.

To address this issue, a suspended EBG is proposed here, as shown
in Fig. \ref{single sub-6GHz element} (b) and Fig. \ref{Orthographic views single mm-wave element}
(d) and (e). Instead of directly connecting to the ground, the vias
of the suspended EBG are connected to a long metallic strip at the
back of substrate 2 that is above the ground and separated by a thin
bondply (0.204 mm), as shown in Fig. \ref{cross view EBG}. The strip
and the ground can then work as a capacitor, coupling the signal from
vias to the ground. By adjusting the dimensions of the strip, the
suspended EBG can work similarly to the traditional EBG. Fig. \ref{surface current distribution}
(c) shows that the surface wave is well suppressed after the suspended
EBG is introduced. The return loss and radiation pattern also demonstrate
good performance, as shown in Fig. \ref{performance of different EBG}.

\subsection{$8\times8$ mmWave RIS}

\begin{figure}[t]
\begin{centering}
\textsf{\includegraphics[height=0.45\columnwidth]{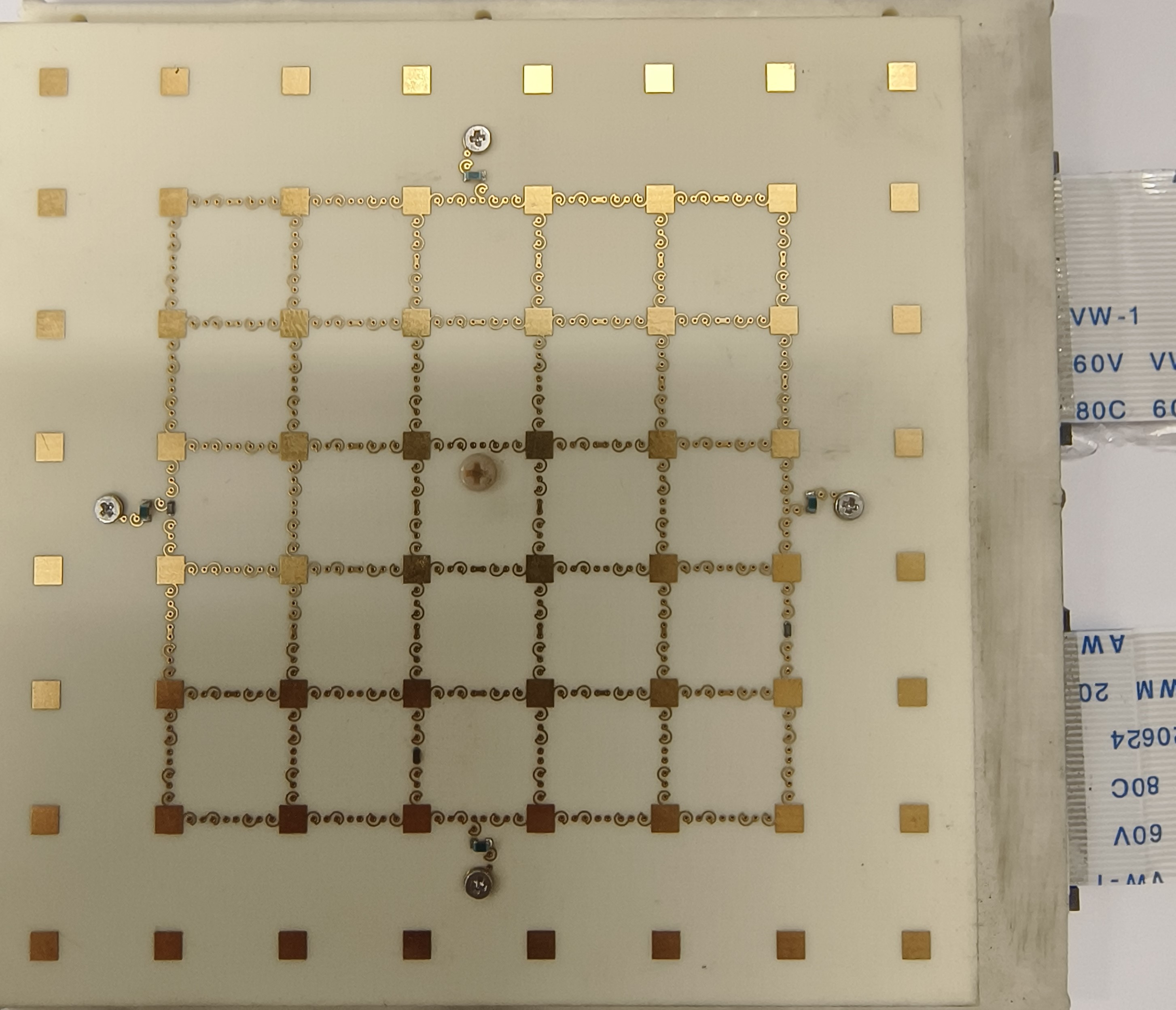}\includegraphics[height=0.45\columnwidth]{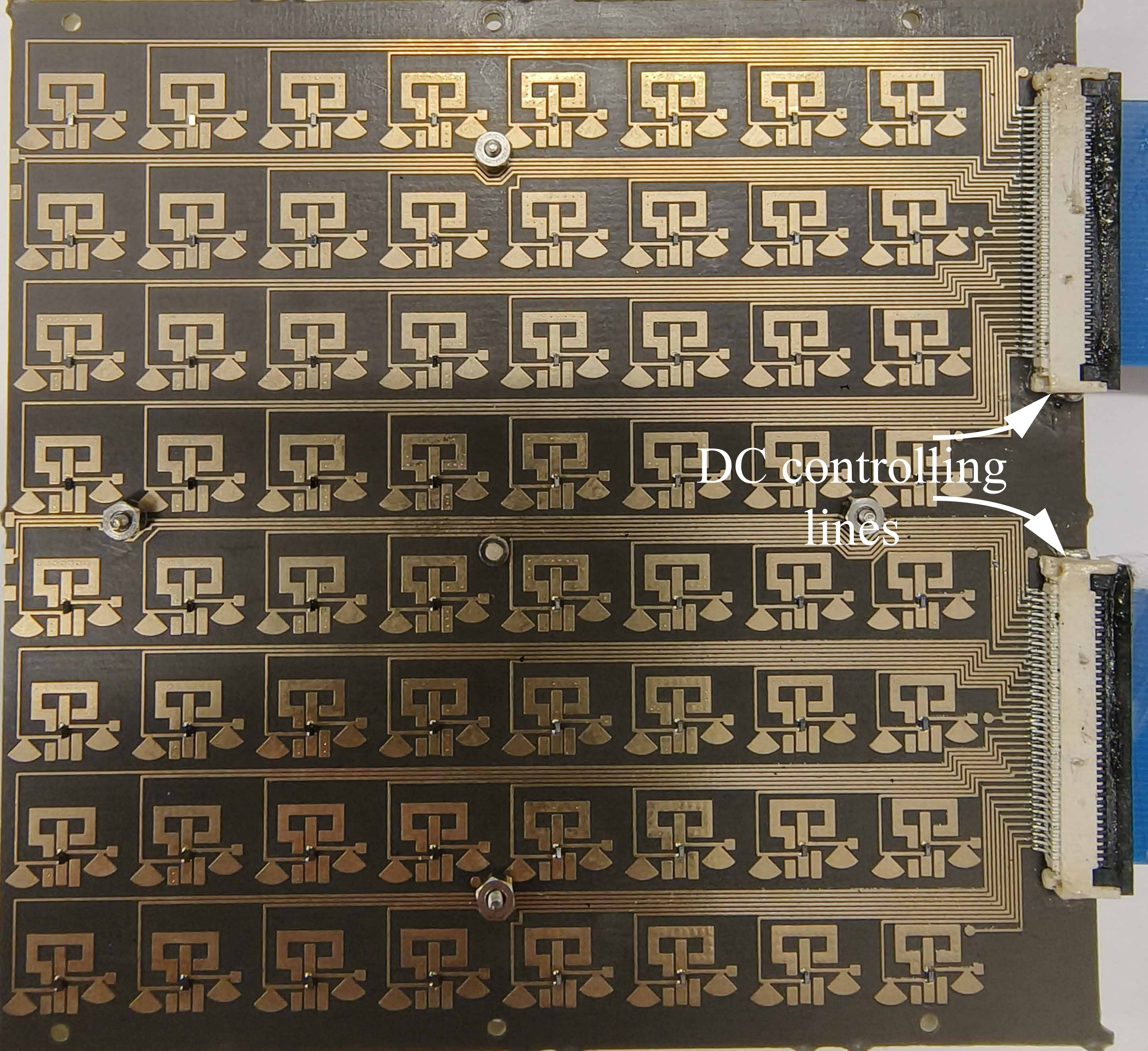}}
\par\end{centering}
\begin{raggedright}
\hspace*{0.2\columnwidth}(a)\hspace*{0.44\columnwidth} (b)
\par\end{raggedright}
\caption{Photograph of the fabricated single sub-6 GHz element consisting of
8$\times$8 mmWave elements with designed DC controlling lines. (a)
Top view (upper patch layer). (b) Bottom view (mmWave circuit layer).}
\label{picture 8x8 mmwave ris}
\end{figure}

By combining the mmWave antenna and phase shifter, a 1-bit mmWave
element can be obtained. We therefore formed and fabricated a DBI-RIS
containing $8\times8$ mmWave elements, as shown in Fig. \ref{picture 8x8 mmwave ris},
where the DC controlling network for it is carefully routed in the
mmWave circuit layer. This $8\times8$ mmWave RIS is also used later
as a single sub-6 GHz element. To analytically calculate the scattered
field by the fabricated DBI-RIS, a method based on the theory of Th�venin
equivalent circuits and the linear superposition of electromagnetic
waves in free space can be utilized, which is detailed in \cite{Rao2023}.

The scattered field pattern excited by an external incident wave can
be found using
\begin{equation}
E_{s}\left(\Omega\right)=\sum_{m=1}^{M}i_{m}E_{m}\left(\Omega\right)+E_{\mathrm{oc}}\left(\Omega\right),\label{scattered pattern}
\end{equation}
where $E_{\mathrm{oc}}\left(\Omega\right)$ is the scattered pattern
when all the mmWave antennas are open-circuited and $E_{m}\left(\theta,\varphi\right)$
is the electric fields radiated by the $8\times8$ mmWave RIS when
a unit current source is excited at the port of the $m$th antenna
with all the other antennas open-circuited. The current $i_{m}$ $\forall m$
is the current excited by incident wave through the $m$th antenna
port for a specific combination of the states at each element.

In the fabricated RIS, each mmWave element has two states that can
provide two scattered fields with different phases. Therefore, for
an mmWave RIS with 8$\mathbf{\times}$8 elements, a total of $2^{64}$
combinations of states are available. Since the calculation of \eqref{scattered pattern}
is analytical, it is much faster than full-wave simulation software,
such as CST \cite{Song2014}. This provides an easy path to optimizing
the scattered patterns in the desired directions.

\begin{figure}[t]
\begin{centering}
\textsf{\includegraphics[width=0.9\columnwidth]{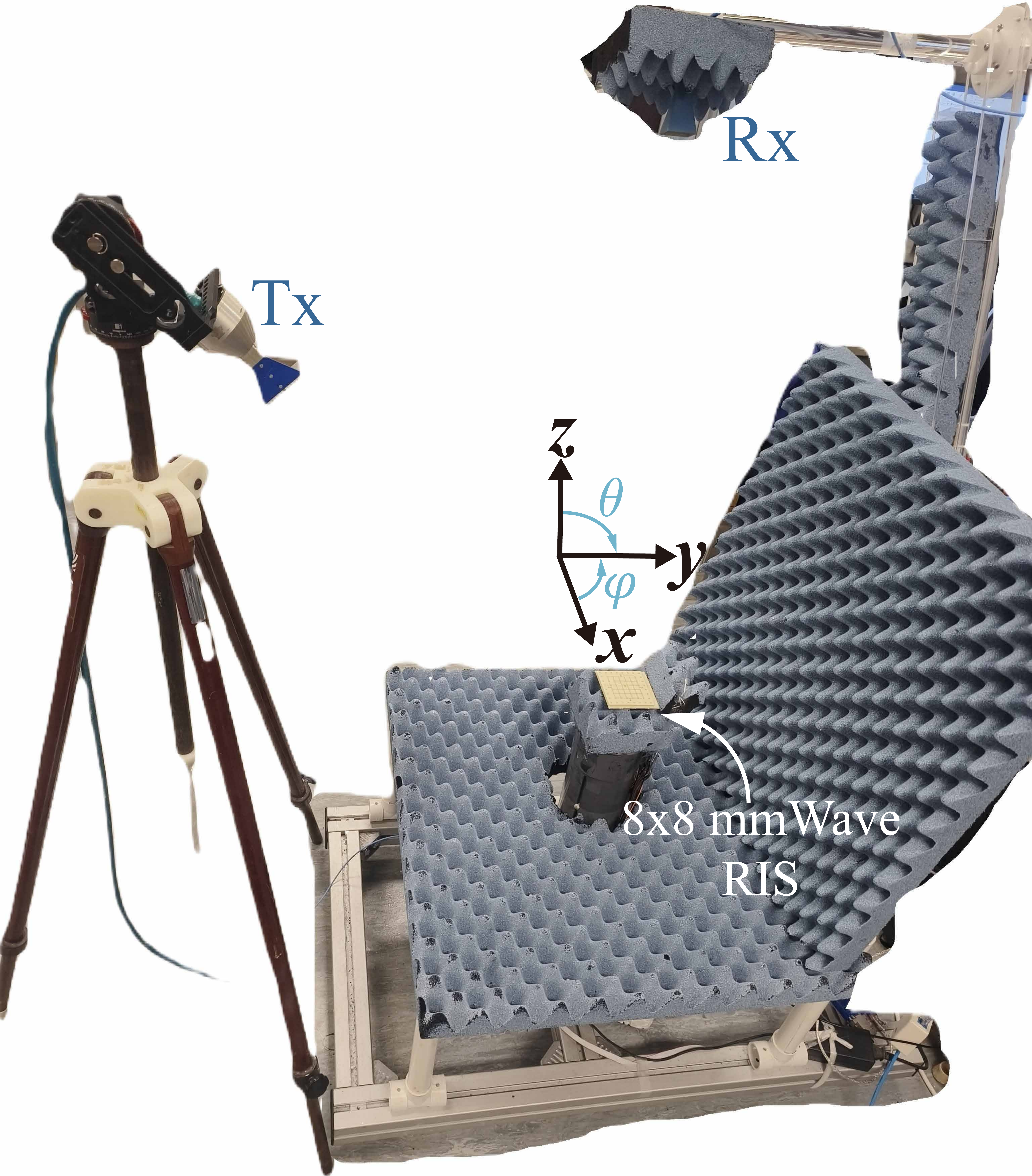}}
\par\end{centering}
\caption{Photograph of the experiment setup for measuring the mmWave scattered
wave pattern of the DBI-RIS with 8$\times$8 mmWave elements. The
configuration of the coordinate system is also shown.}
\label{Measurement setup mmwave RIS}
\end{figure}

A measurement setup is also designed and built, as shown in Fig. \ref{Measurement setup mmwave RIS},
where the transmit (Tx) antenna is fixed with a specific incident
angle, and the receive (Rx) antenna can rotate over the $xoz$ plane
to measure the scattered pattern. The $8\times8$ mmWave RIS is fixed
in the testbed, while the DC controlling lines and FPGA are shielded
by the absorber.

\begin{figure}[t]
\begin{centering}
\textsf{\includegraphics[width=0.33\columnwidth]{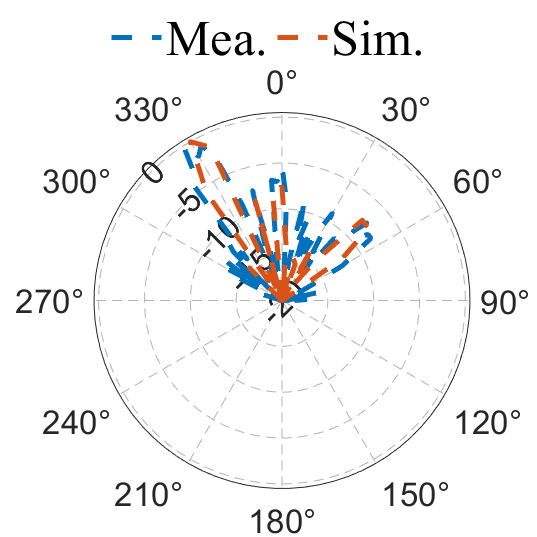}\includegraphics[width=0.33\columnwidth]{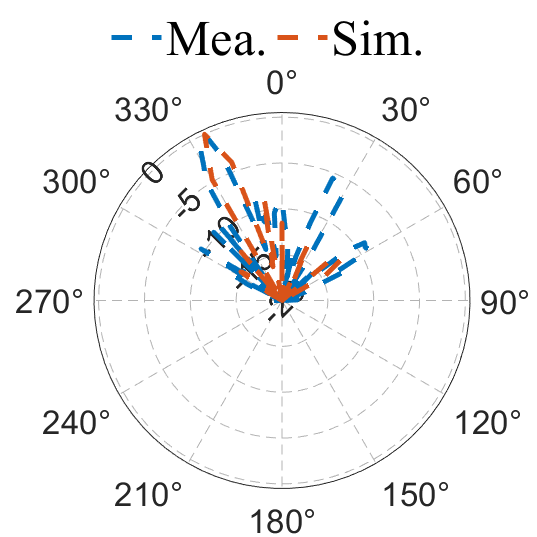}\includegraphics[width=0.33\columnwidth]{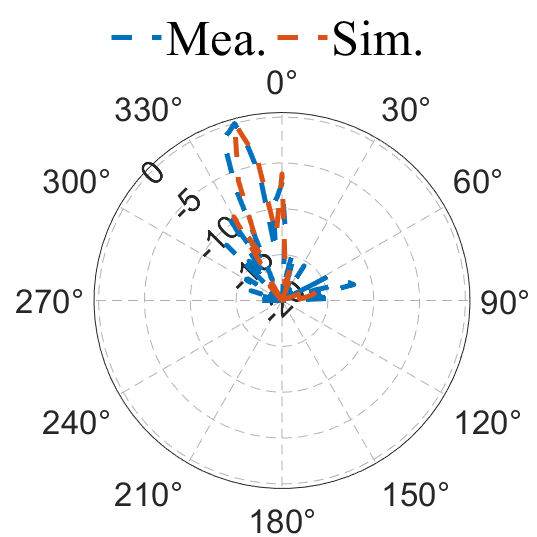}}
\par\end{centering}
\begin{raggedright}
\hspace*{0.15\columnwidth}(a)\hspace*{0.27\columnwidth} (b)\hspace*{0.27\columnwidth}
(c)
\par\end{raggedright}
\begin{centering}
\textsf{\includegraphics[width=0.33\columnwidth]{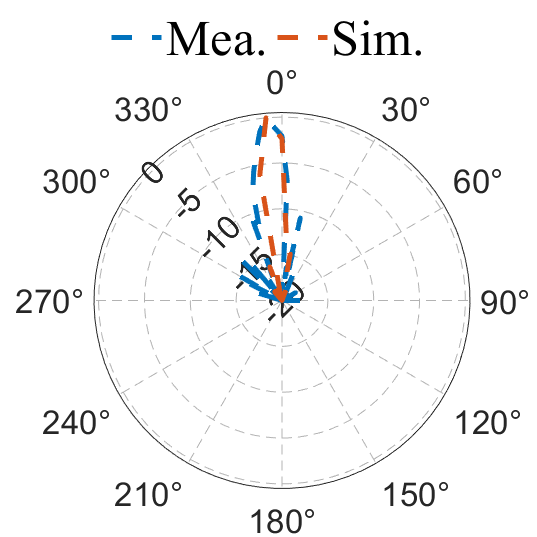}\includegraphics[width=0.33\columnwidth]{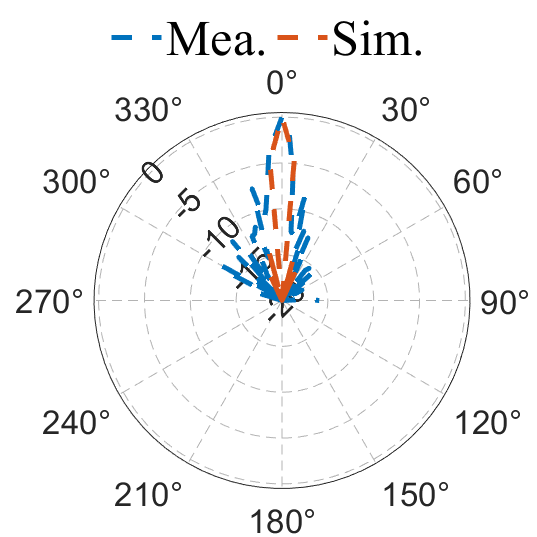}\includegraphics[width=0.33\columnwidth]{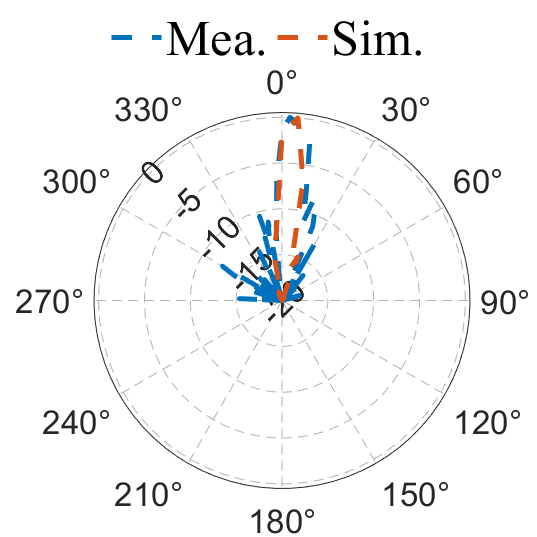}}
\par\end{centering}
\begin{raggedright}
\hspace*{0.15\columnwidth}(d)\hspace*{0.27\columnwidth} (e)\hspace*{0.27\columnwidth}
(f)
\par\end{raggedright}
\begin{centering}
\textsf{\includegraphics[width=0.33\columnwidth]{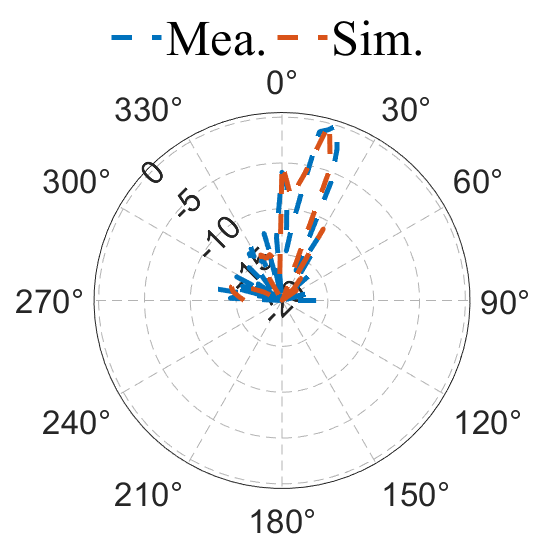}\includegraphics[width=0.33\columnwidth]{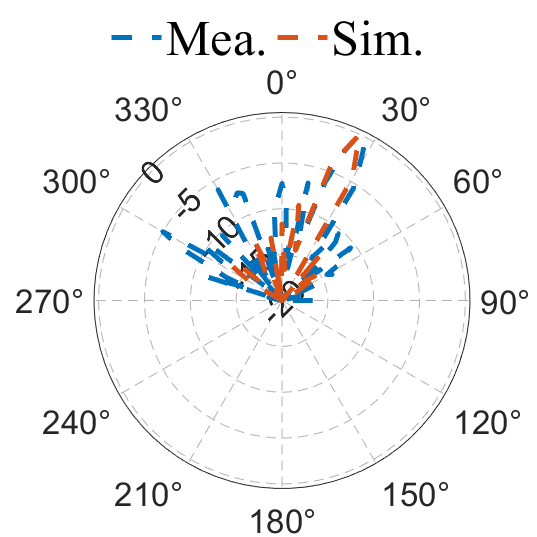}\includegraphics[width=0.33\columnwidth]{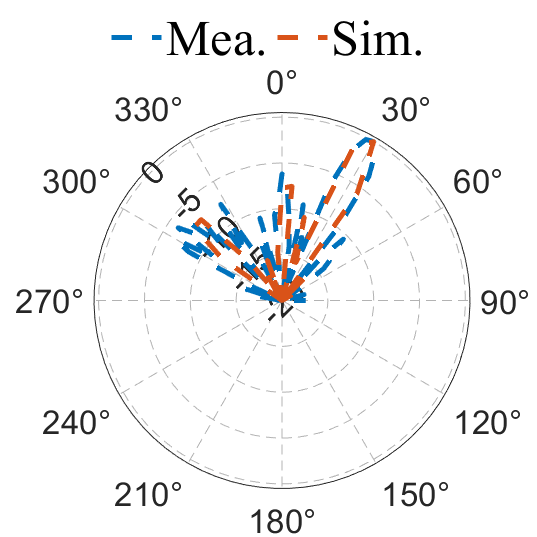}}
\par\end{centering}
\begin{raggedright}
\hspace*{0.15\columnwidth}(g)\hspace*{0.27\columnwidth} (h)\hspace*{0.27\columnwidth}
(i)
\par\end{raggedright}
\caption{Nine examples of the simulated and measured scattered pattern for
28 GHz when the incident angle $(\theta,\phi)_{\mathrm{incident}}=(45^{\circ},270^{\circ})$,
and the main beam is steered to (a) $(\theta,\phi)_{\mathrm{beam}}=(-30^{\circ},0^{\circ})$,
(b) $(\theta,\phi)_{\mathrm{beam}}=(-25^{\circ},0^{\circ})$, (c)
$(\theta,\phi)_{\mathrm{beam}}=(-15^{\circ},0^{\circ})$, (d) $(\theta,\phi)_{\mathrm{beam}}=(-5^{\circ},0^{\circ})$,
(e) $(\theta,\phi)_{\mathrm{beam}}=(0^{\circ},0^{\circ})$, (f) $(\theta,\phi)_{\mathrm{beam}}=(5^{\circ},0^{\circ})$,
(g) $(\theta,\phi)_{\mathrm{beam}}=(15^{\circ},0^{\circ})$, (h) $(\theta,\phi)_{\mathrm{beam}}=(25^{\circ},0^{\circ})$,
(i) $(\theta,\phi)_{\mathrm{beam}}=(30^{\circ},0^{\circ})$.}
\label{mea and sim scattered pattern of 8x8 mmwave RIS}
\end{figure}

\begin{figure}[t]
\begin{centering}
\textsf{\includegraphics[width=0.33\columnwidth]{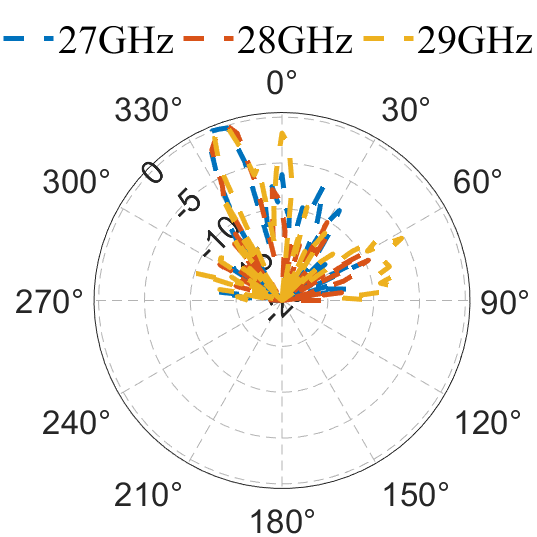}\includegraphics[width=0.33\columnwidth]{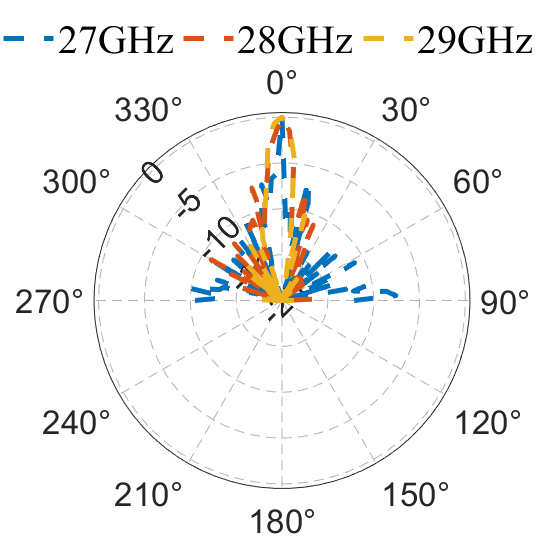}\includegraphics[width=0.33\columnwidth]{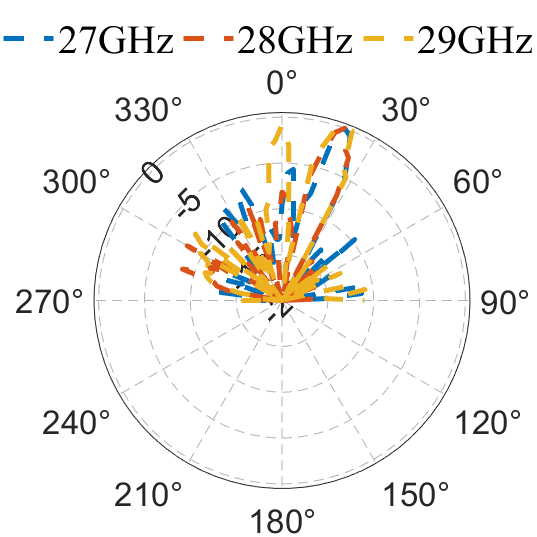}}
\par\end{centering}
\begin{raggedright}
\hspace*{0.15\columnwidth}(a)\hspace*{0.27\columnwidth} (b)\hspace*{0.27\columnwidth}
(c)
\par\end{raggedright}
\caption{The measured scattered pattern for different frequencies and the main
beam is steered to (a) $(\theta,\phi)_{\mathrm{beam}}=(-20^{\circ},0^{\circ})$,
(b) $(\theta,\phi)_{\mathrm{beam}}=(0^{\circ},0^{\circ})$, (c) $(\theta,\phi)_{\mathrm{beam}}=(20^{\circ},0^{\circ})$.}
\label{mea scattered pattern of 8x8 mmwave RIS different fres}
\end{figure}

We fixed the incident angle at $45^{\circ}$, and the comparison between
simulated and measured scattered patterns with different main directions
is demonstrated in Fig. \ref{mea and sim scattered pattern of 8x8 mmwave RIS}.
The scattered main beam can be successfully steered between $-30^{\circ}$
to $30^{\circ}$ with good alignment with simulated results. Fig.
\ref{mea scattered pattern of 8x8 mmwave RIS different fres} demonstrates
the measured scattered pattern at different frequencies, which shows
the fabricated RIS can well cover the bandwidth of 27-29 GHz, except
for a slightly larger side lobe at $0^{\circ}$ for 29 GHz due to
the slightly poor matched performance of the mmWave antenna, as shown
in Fig. \ref{performance of mm-wave antenna} (a). 

\section{Sub-6 GHz RIS Working Methodology}

\subsection{Architecture of a single sub-6 GHz element}

\begin{figure}[t]
\begin{centering}
\textsf{\includegraphics[width=0.85\columnwidth]{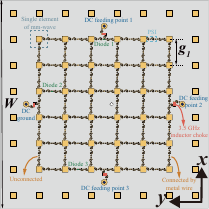}}
\par\end{centering}
\caption{Geometry of the proposed sub-6 GHz element. The central 6$\times$6
mmWave patches (out of the 8$\times$8 mmWave patches) are selectively
connected with three RF switches to provide reconfigurability. The
dimensions are $g_{1}=8.5$, $W=68$. (Unit: mm)}
\label{Sub-6 GHz element geometry}
\end{figure}

To realize the function of a sub-6 GHz RIS element, the 8$\times$8
mmWave RIS is reused as a single element for the sub-6 GHz band based
on a shared-aperture structure as shown in Fig. \ref{Sub-6 GHz element geometry}.
The central 6$\times$6 mmWave elements are connected together using
PSIs that function as chokes to prevent short circuit connections
between the mmWave antennas. The outer edge elements of the 8$\times$8
mmWave elements are left unconnected. Reconfigurability is achieved
using three switches constructed from PIN diodes among the central
6$\times$6 mmWave elements. The number of diodes in a single sub-6
GHz element is set to 3 to balance the trade-off between complexity
and reconfigurability. In addition, to reduce the complexity of the
controlling network, the cathodes of all diodes are connected together
so that there are a total of only four DC feeding points in a single
sub-6 GHz element. All of the DC feeding points are isolated from
the main structure by DC chokes and led to the back of the ground
to \textquotedbl mmWave circuit layer\textquotedbl{} by four metal
screws, as shown in Fig. \ref{picture 8x8 mmwave ris}. These screws
are also used for fixing the entire element,

\subsection{Planar spiral inductor (PSI)}

The dimension of a single mmWave element is much smaller than a wavelength
at sub-6 GHz; thus, an array of mmWave elements must be connected
together appropriately to form a larger structure for the sub-6 GHz
element. However, direct connection using metal wires is not a good
choice since it will alter the main structure of the mmWave element
and destroy its effectiveness. One possible solution is to use commercial
inductor chokes at the mmWave band to isolate the connecting metal
wires from the mmWave parch. However, the cost of a single inductor
choke at such a high frequency is high, and the number of inductor
needed is also very large due to the large number of mmWave elements,
leading to high costs and impracticality. Therefore, PSI is proposed
in this paper for connecting mmWave elements together that can isolate
mmWave signals and pass through sub-6 GHz signals. Additionally, PSI
is a structure in the substrate and fabricated along with the PCB
process, providing a low-cost method.

\begin{figure}[t]
\begin{centering}
\textsf{\includegraphics[width=0.7\columnwidth]{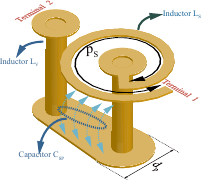}}
\par\end{centering}
\begin{centering}
(a)
\par\end{centering}
\begin{centering}
\textsf{\includegraphics[width=0.85\columnwidth]{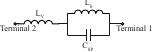}}
\par\end{centering}
\begin{centering}
(b)
\par\end{centering}
\caption{(a) The proposed PSI with noted key geometric parameters and equivalent
components (b) Schematic of the equivalent circuit for PSI. }
\label{proposed PSI}
\end{figure}

The PSI structure is shown in Fig. \ref{proposed PSI} (a), which
has two terminals. Terminal 1 is at the endpoint of a spiral ring,
while the other endpoint is connected to a metallic bottom plate by
a via. Terminal 2 is also connected to the bottom plate by another
via. To analyze this PSI, the spiral ring can first be thought of
as an inductor, $\mathrm{L_{S}}$, since its perimeter is comparable
to a wavelength at the mmWave band. Then, the spiral ring will also
couple the signal from itself to the bottom plate and can be equivalent
to a capacitor, $\mathrm{C_{SP}}$, that is in parallel to $\mathrm{L_{S}}$.
Finally, the via of terminal 2 can also be treated as an inductor,
$\mathrm{L_{V}}$, that is in series with the former parallel LC circuit,
as shown in Fig. \ref{proposed PSI} (b).

\begin{figure}[t]
\begin{centering}
\textsf{\includegraphics[width=0.5\columnwidth]{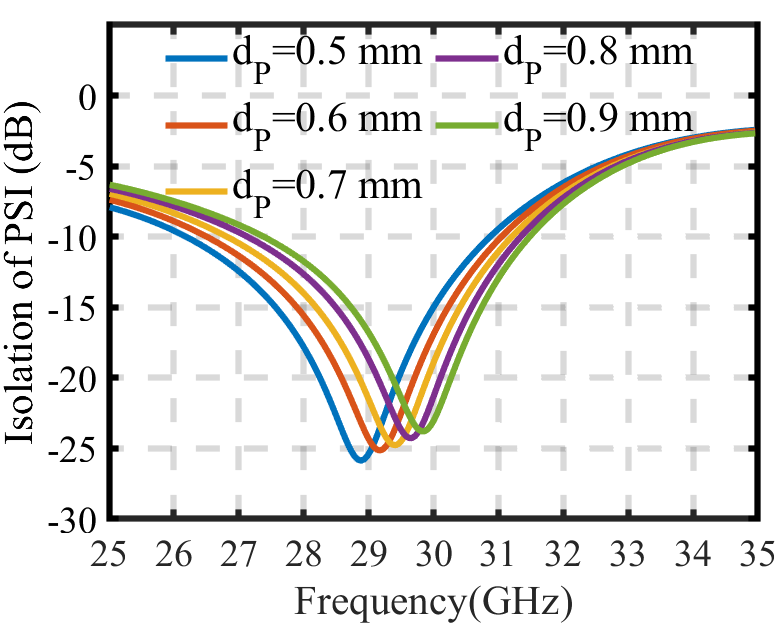}\includegraphics[width=0.5\columnwidth]{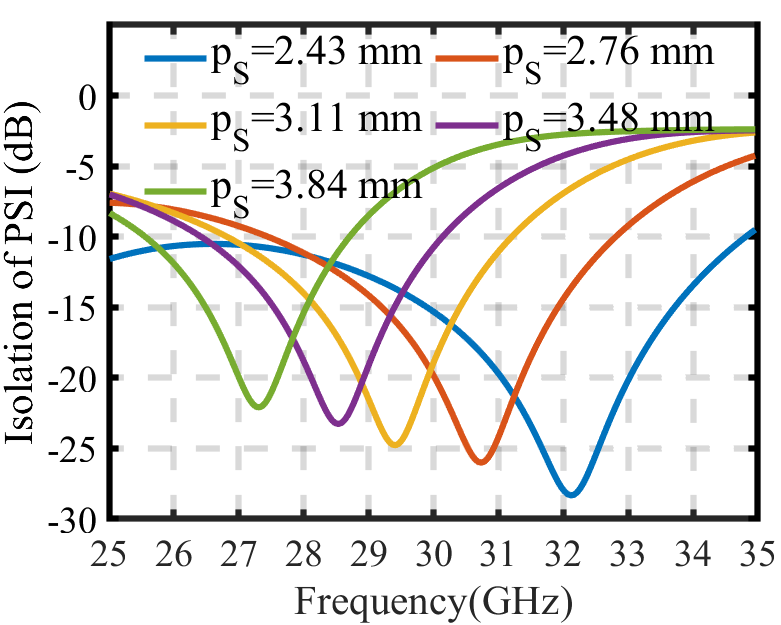}}
\par\end{centering}
\begin{raggedright}
\hspace*{0.24\columnwidth}(a)\hspace*{0.44\columnwidth} (b)
\par\end{raggedright}
\caption{The simulated isolation of the proposed PSI for mmWave band with (a)
different $\mathrm{d}_{\mathrm{p}}$ and $\mathrm{p}_{\mathrm{S}}=3.11$
mm and (b) different $\mathrm{p}_{\mathrm{S}}$ and $\mathrm{d}_{\mathrm{p}}=0.7$
mm. }
\label{PSI parameter sweeping}
\end{figure}

The primary requirement for PSI is to provide high isolation between
the two terminals at the mmWave band. To achieve this, the parallel
LC, $\mathrm{C_{SP}}$ and $\mathrm{L_{S}}$, can be designed to resonate
at the target frequency, 28 GHz in this paper. The resonating frequency
is $1/(2\pi\sqrt{C_{SP}\mathrm{L_{S}}})$ and can be tuned by adjusting
the values of $\mathrm{C_{SP}}$ and $\mathrm{L_{S}}$, which are
determined by the spiral ring perimeter, $\mathrm{p_{s}}$, and diameter
of the bottom plate, $\mathrm{d_{P}}$. Fig. \ref{PSI parameter sweeping}
(a) demonstrates the isolation of PSI for different $\mathrm{d_{P}}$.
By increasing $\mathrm{d_{P}}$, the resonating frequency can be tuned
to higher values. Fig. \ref{PSI parameter sweeping} (b) shows the
isolation of PSI for different $\mathrm{P_{S}}$. Increasing $\mathrm{P_{S}}$
will enlarge the inductor $\mathrm{L_{S}}$ and thus lead to a lower
resonating frequency.

\begin{figure}[t]
\begin{centering}
\textsf{\includegraphics[width=0.5\columnwidth]{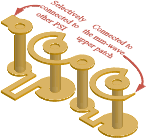}\includegraphics[width=0.5\columnwidth]{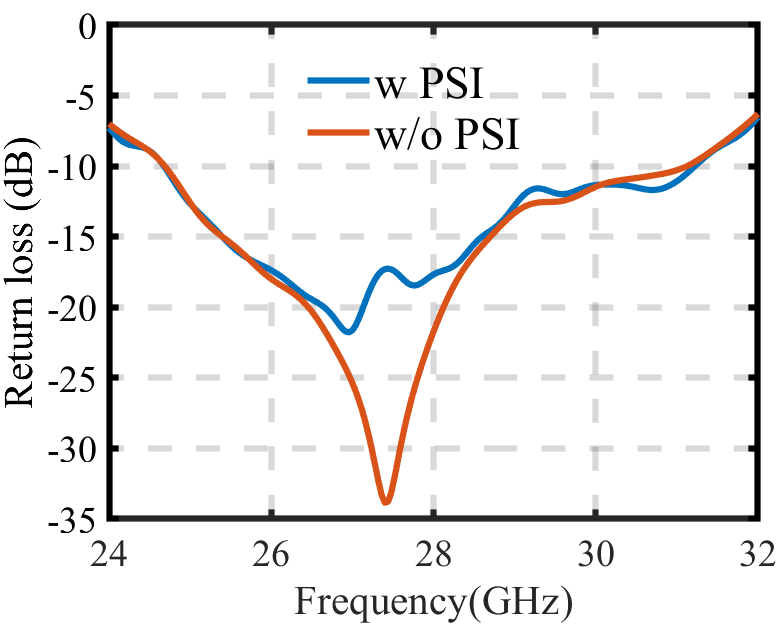}}
\par\end{centering}
\begin{raggedright}
\hspace*{0.24\columnwidth}(a)\hspace*{0.44\columnwidth} (b)
\par\end{raggedright}
\caption{(a) Two PSIs are cascaded together with folded line at the bottom
to be applied in the proposed DBI-RIS. (b) Simulated performance of
the proposed mmWave antenna with and without PSI. }
\label{cascaded two PSI}
\end{figure}

For PSI that resonate at the designed frequency, high isolation is
achieved for the two terminals, but the PSI structure itself will
also have a near-field effect on the performance of the mmWave antenna.
The length of the bottom plate is close to the length of the mmWave
element upper patch. Therefore, the PSI in Fig. \ref{proposed PSI}
(a) cannot be directly used in our DBI-RIS; otherwise, its bottom
plate will also radiate and significantly affect the performance of
the mmWave antenna. Thus, we changed the bottom plate to a folded
line, as shown in Fig. \ref{cascaded two PSI} (a), and two PSIs with
different lengths of folded lines are also cascaded to further avoid
the near-field effect on the mmWave antenna. Fig. \ref{cascaded two PSI}
(b) shows the return loss of the antenna with and without the cascaded
PSI in the four sides of the upper patch. The high isolation and folded
bottom lines enable the PSI to be almost invisible to the mmWave antenna.
In addition, since the dimension of the PSI is much smaller than a
wavelength of the sub-6 GHz band, it works similarly to a metal wire
in this band. These properties make it suitable as a connecting structure
among the mmWave elements, facilitating the formation of the sub-6
GHz element.

\subsection{Optimization of the sub-6 GHz element}

\begin{figure}[t]
\begin{centering}
\textsf{\includegraphics[width=0.85\columnwidth]{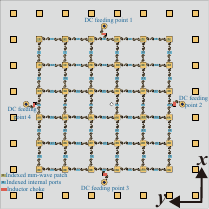}}
\par\end{centering}
\caption{The original sub-6 GHz element formed by 8$\times$8 mmWave elements.
For the central 6$\times$6 mmWave patches, the gaps between adjacent
PSI can be modeled as internal ports allowing the structure to be
analyzed using a multi-port model. And the central 6$\times$6 mmWave
patches are also numbered for the analysis of RF switches' DC feedings.}
\label{multi_pots_scematic_of_3_5GHz_element}
\end{figure}

With the PSI, the mmWave elements can be connected together without
affecting the performance of the mmWave band. To realize the reconfigurability
function in the sub-6 GHz band by reusing the mmWave element, three
requirements need to be fulfilled, including effectiveness over a
broad angular range, sufficient bandwidth, and electronically reconfigurable
phase shifts to the incident waves. The proposed sub-6 GHz element
consists of an $8\times8$ mmWave element array with the central $6\times6$
elements used for the main structure. The outer elements ensure low
mutual coupling between adjacent sub-6 GHz elements, as shown in Fig.
\ref{multi_pots_scematic_of_3_5GHz_element}. Since all the mmWave
elements are small compared to a wavelength of the sub-6 GHz band,
the mmWave element array without any connections will have a negligible
effect on the sub-6 GHz incident wave and be nearly invisible to it.
However, by optimally selecting the connections between mmWave elements
and appropriately placing RF switches between some adjacent mmWave
elements, the array can be utilized to realize the sub-6 GHz element.
This leverages the single band approach detailed in \cite{Rao2022},
which will be briefly introduced in this paper for convenience. However,
since the DC feeding points are kept to a minimum number and have
fixed positions, an extra constraint compared to the method in \cite{Rao2022}
needs to be applied.

In the proposed sub-6 GHz element design, the complexity is highly
related to the number of RF switches in a single element since more
RF switches require more DC controlling lines and introduce more undesired
effects in the operation of the mmWave RIS. As a balance between the
complexity and reconfigurability of the sub-6 GHz element, we set
the number of RF switches $Q=3$, meaning that each sub-6 GHz element
can provide 8 different reconfigurable states with distinct scattering
characteristics to the incident electromagnetic waves. Instead of
optimizing the reflected phases of all 8 states one by one, we first
define a unified metric, phase entropy \cite{Rao2022}, to characterize
the performance of the sub-6 GHz element. We assume the reflection
phases $\varphi_{i}$ for $i=1,2,...,2^{Q}$ are sorted in an ascending
order, that is $\varphi_{i}\geq\varphi_{j}$ if $i\geq j$, and we
use degrees as the unit of phase, that is $0\leqslant\varphi_{i}<360^{\circ}$
$\forall i$. Then, the difference between adjacent phases is 
\begin{equation}
\begin{cases}
\Delta\varphi_{i}=\varphi_{i+1}-\varphi_{i}, & 1\leqslant i<2^{Q}\\
\Delta\varphi_{i}=2\pi+\varphi_{1}-\varphi_{i}, & i=2^{Q}
\end{cases},\label{eq:2}
\end{equation}
where $\Delta\varphi_{i}$ for $i=1,2,...,2^{Q}$ are positive and
satisfy that $\sum_{i=1}^{2^{Q}}\Delta\varphi_{i}=360^{\circ}$.

Phase entropy is then defined as
\begin{equation}
H=-\sum_{i=1}^{2^{Q}}\left(\frac{\Delta\varphi_{i}}{360^{\circ}}\log_{2}\left(\frac{\Delta\varphi_{i}}{360^{\circ}}\right)\right),\label{eq:3}
\end{equation}
where $H$ is a function of the positions of RF switches and connection
configuration among the mmWave elements. The maximum value of $H$
for $Q=3$ switches is 3 and the bigger $H$ means the better performance
of sub-6 GHz element and the better reconfigurability.

In order to obtain  $H$, we will need to know the scattered pattern
for all 8 states of the sub-6 GHz element. If the gaps between PSIs
are treated as internal ports (internal to the DBI-RIS structure)
as shown in Fig. \ref{multi_pots_scematic_of_3_5GHz_element}, based
on the linear superposition of electromagnetic waves in free space
\cite{Balanis2012} \cite{Rao2022}, the scattered wave from the sub-6
GHz element can be written as
\begin{equation}
E_{s}\left(\Omega\right)=\sum_{m=1}^{M}i_{m}E_{m}\left(\Omega\right)+E_{\mathrm{oc}}\left(\Omega\right),\label{eq4}
\end{equation}
where $E_{\mathrm{oc}}\left(\Omega\right)$ is the scattered wave
pattern when all internal ports are open-circuited and $E_{m}\left(\theta,\varphi\right)$
is the electric fields radiated by the RIS element when a unit current
source is connected to the $m$th internal port with all the other
ports open-circuit and no waves incident. The currents $i_{m}$ $\forall m$
is the current through the $m$th internal port and can be found as{\small{}
\begin{equation}
\mathbf{i}=-\left(\mathbf{Z}+\mathbf{Z}_{L}\right)^{-1}\mathbf{v}_{\mathrm{oc}},\label{eq5}
\end{equation}
where }diagonal matrix $\mathbf{Z}_{L}$ is the equivalent load at
the internal ports, representing the connecting status of all internal
ports, $\mathbf{v}_{\mathrm{oc}}=\left[v_{\mathrm{oc},1},v_{\mathrm{oc},2},\ldots,v_{\mathrm{oc},M}\right]^{T}$
with $v_{\mathrm{oc},m}$ denoting the open-circuit voltage induced
at the $m$th internal port and $\mathbf{Z}$ denotes the impedance
matrix of all the internal ports. The detailed calculation procedures
can be found in \cite{Rao2022}. 
\begin{figure}[t]
\begin{centering}
\textsf{\includegraphics[width=0.5\columnwidth]{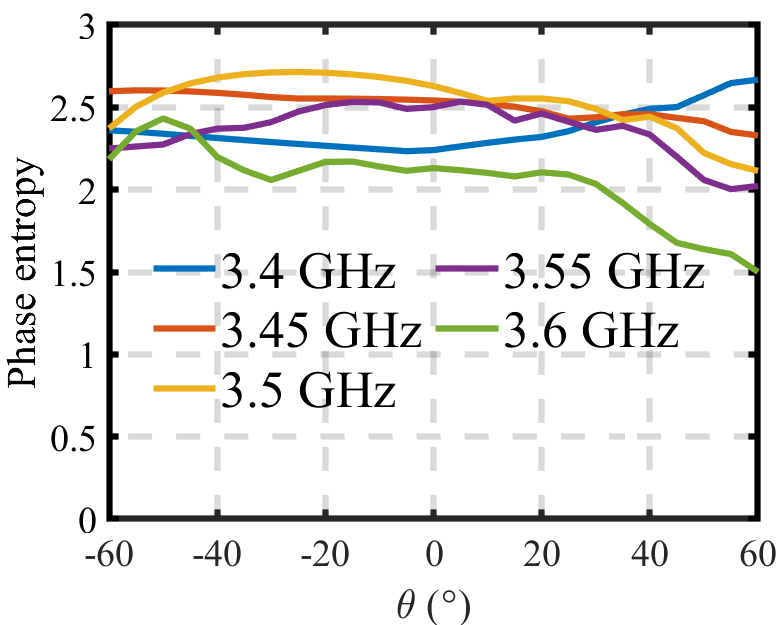}\includegraphics[width=0.5\columnwidth]{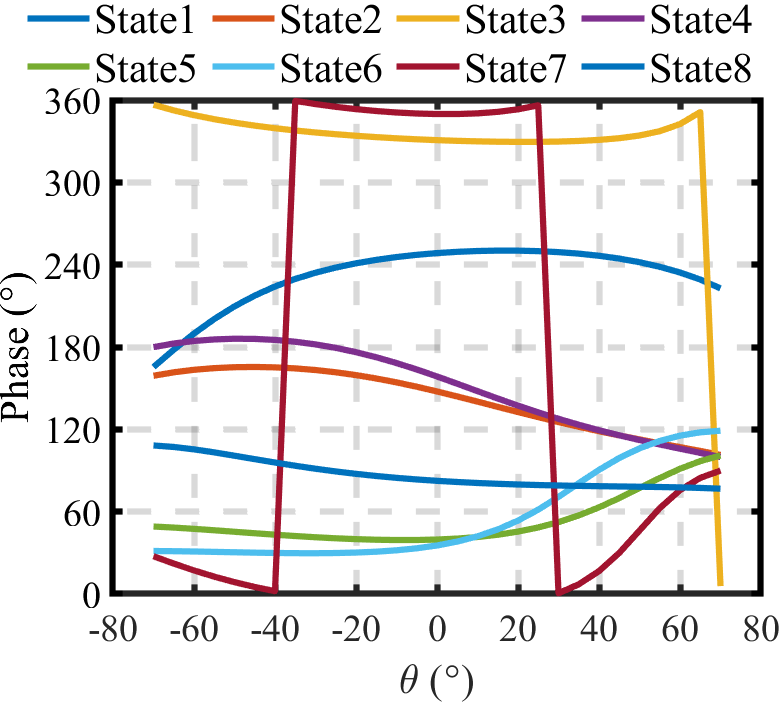}}
\par\end{centering}
\begin{raggedright}
\hspace*{0.24\columnwidth}(a)\hspace*{0.44\columnwidth} (b)
\par\end{raggedright}
\caption{Performance of sub-6 GHz element when it is excited by vertically
incident wave. (a) Phase entropy versus $\theta$ for different frequencies
(at 3.4, 3.45, 3.5, 3.55 and 3.6 GHz) are shown. (b) Reflection phases
are shown versus $\theta$ for 8 different states of sub-6 GHz element
at 3.5 GHz.}
\label{Performance of sub-6 GHz element}
\end{figure}

There are a total of $M=60$ internal ports for the sub-6 GHz element.
To represent the connection state of each internal port, we use binary
variables $x_{m}\in\left\{ 0,1\right\} $, where $m$ ranges from
1 to $M$. A value of 0 denotes an unconnected state, while 1 represents
a connected state for the $m$th internal port. The geometry of a
specific configuration of connections within the sub-6 GHz element
is represented by a binary connection vector $\mathbf{x_{\mathrm{0}}}=\left[x_{1},x_{2},\ldots,x_{M}\right]$.
To capture the reconfigurability of $Q=3$ RF switches among internal
ports, we expand $\mathbf{x_{\mathrm{0}}}$ by adding the positions
of the 3 reconfigurable switches. This expansion yields a geometry
vector $\mathbf{x}=\left[\mathbf{x}_{0},\mathbf{x}_{1},\ldots,\mathbf{x}_{Q}\right]$,
where $\mathbf{x}_{q}$ for $q=1,2,...,Q$ is a $\left\lceil \log_{2}M\right\rceil $
(where $\left\lceil \cdot\right\rceil $ denotes the ceiling function)
binary vector, representing the positions of the $q$th RF switch.
Therefore, we can find that $H$ is a function of vector $\mathbf{x}$,
denoted $H(\mathbf{x})$. With \eqref{eq:3} and \eqref{eq4}, we
can find the analytical expression of $H(\mathbf{x})$ after a single
full-wave simulation to obtain $\mathbf{Z},$ $\mathbf{Z}_{L}$ and
$\mathbf{v}_{\mathrm{oc}}$, which makes the computation very efficient.
This enables us to optimize the connecting states among the mmWave
elements and the positions of RF switches using mature algorithms
such as the genetic algorithm (GA).

Besides the calculation of $H(\mathbf{x})$, \eqref{eq:3}, one special
constraint is also required to be applied in the design of the sub-6
GHz element. For the three RF switches, we have limited the number
of DC controlling lines to a minimum of four, with a shared ground
and three independent controlling lines. In addition, to avoid the
influence of DC lines on the effectiveness of mmWave elements, we
also fixed their positions at the edges of the mmWave element, as
shown in Fig. \ref{multi_pots_scematic_of_3_5GHz_element}, denoted
as DC feeding points. To ensure that all RF switches can be effectively
controlled, a constraint must fulfill three conditions: 1) the cathodes
of all diodes have to be connected together and lead to one DC feeding
position, 2) the anodes of all diodes have to be connected to the
other three DC feeding points separately, and 3) the four DC feeding
points cannot be electrically connected to each other.

To evaluate these conditions in the optimization process, we can first
index the central $6\times6$ elements as shown in Fig. \ref{multi_pots_scematic_of_3_5GHz_element},
with a total of $N=36$ indexed mmWave elements. The internal ports
and the indexed mmWave element can be fully represented by an $N\times N$
matrix, $\mathbf{Y}$. If the $m$th internal port connects the $n_{1}$th
and $n_{2}$th indexed mmWave elements, $\mathbf{Y}(n_{1}\text{, \ensuremath{n_{2}}})=m$.
On the other hand, for any two $n_{1}$th and $n_{2}$th indexed mmWave
elements, $n_{1}\neq n_{2}$, if they are not linked by an internal
port, $\mathbf{Y}(n_{1}\text{, \ensuremath{n_{2}}})=0$. For any geometry
vector $\mathbf{x}$, by iteratively traversing all $N$ indexed mmWave
elements, we can obtain all the electrically connected indexed mmWave
elements and DC feeding points to any anode or cathode of diodes.
In other words, given the matrix of $\mathbf{Y}$ and $\mathbf{x}$,
the above-mentioned three conditions can be evaluated, and we define
this constraint as a function, $feeding\{\mathbf{Y},\mathbf{x}\}$,
which is 1, meaning the constraint is fulfilled, and 0, meaning the
constraint is unfulfilled.

We can then include this constraint in the optimization of sub-6 GHz
element and formulate it as 

\begin{align}
\mathrm{\underset{\mathbf{\bar{x}}}{max}}\;\; & \frac{1}{KL}\sum_{k=1}^{K}\sum_{l=1}^{L}H\left(\Omega_{k},f_{l},\mathbf{\mathbf{x}}\right)\label{eq:6}\\
\mathrm{s.t.}\;\;\; & \mathbf{x}\in\left\{ 0,1\right\} ^{M+Q\times\left\lceil \log_{2}M\right\rceil }\label{eq:7}\\
 & feeding\{\mathbf{Y},\mathbf{x}\}=1\label{eq:8}
\end{align}
where $\Omega_{k}$ denotes the $k$th spatial angle sample of the
scattered wave for $k=1,...,K$, and $f_{l}$ denotes the $l$th frequency
sample for $l=1,...,L$. In addition we have also updated the entropy
function to explicitly include its dependence on the scattering wave
angle and frequency and write it as $H\left(\Omega_{k},f_{l},\mathbf{\mathbf{x}}\right)$.
The multiple scattered angles and frequencies are considered for ensuring
that the sub-6 GHz can operate over a broad range of angles and sufficient
bandwidth.
\begin{figure}[t]
\begin{centering}
\textsf{\includegraphics[width=0.5\columnwidth]{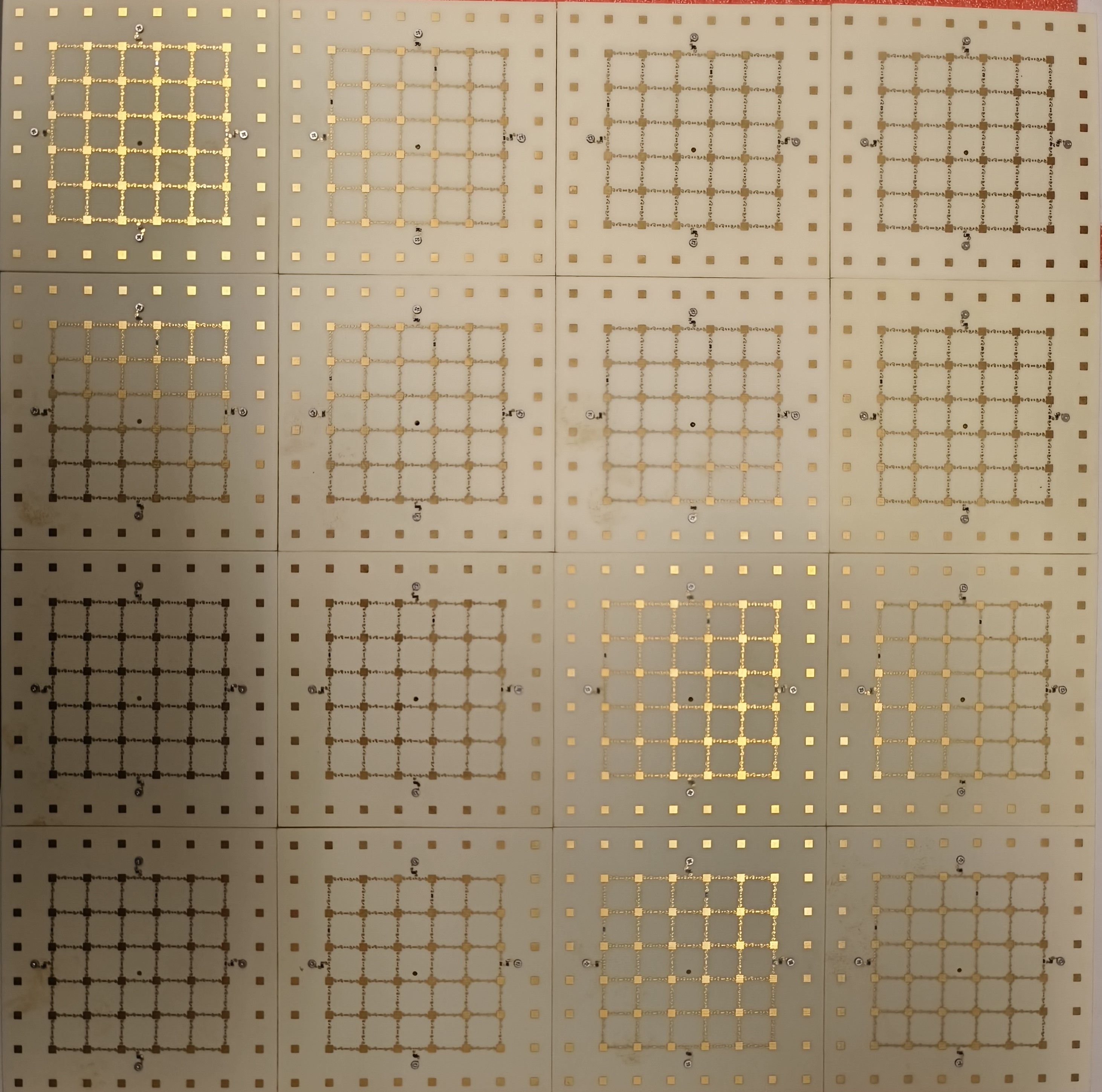}\includegraphics[width=0.5\columnwidth]{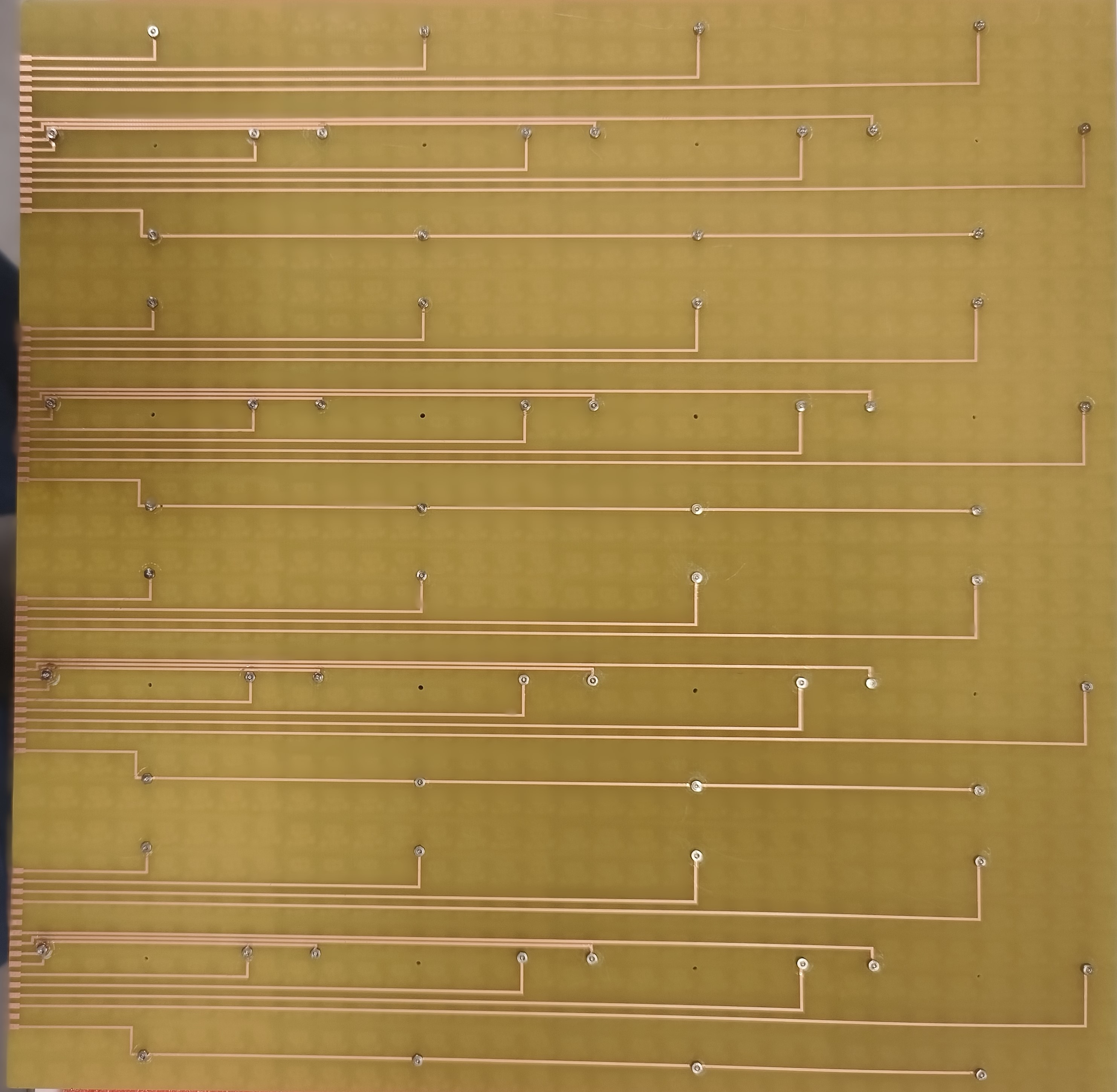}}
\par\end{centering}
\begin{raggedright}
\hspace*{0.24\columnwidth}(a)\hspace*{0.44\columnwidth} (b)
\par\end{raggedright}
\caption{Photographs of the DBI-RIS prototype with 4$\times$4 sub-6 GHz elements.
(a) Top view. (b) Bottom view.}
\label{3_5GHz_4x4_picture}
\end{figure}
\begin{figure}[t]
\begin{centering}
\textsf{\includegraphics[width=0.5\columnwidth]{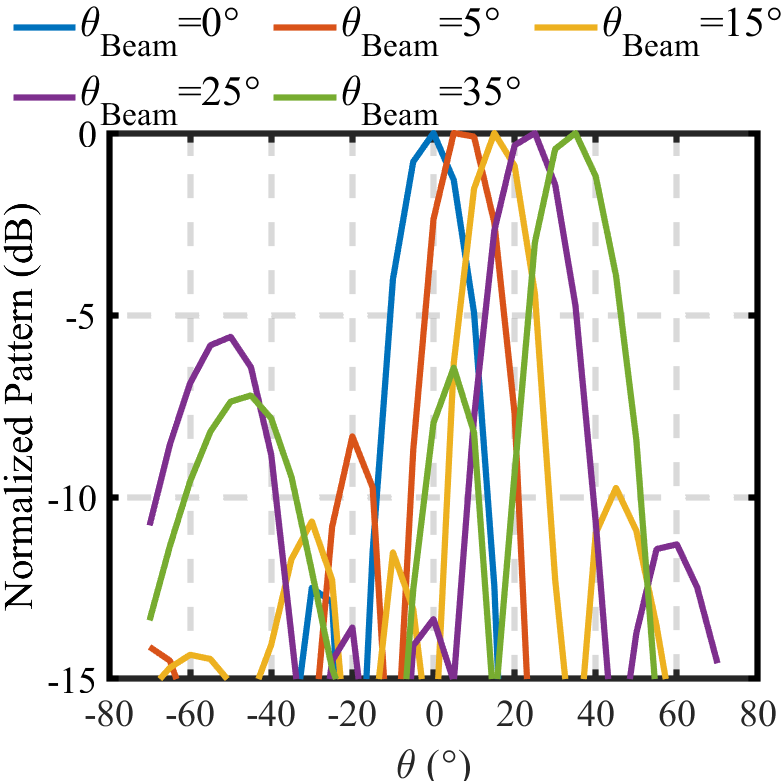}\includegraphics[width=0.5\columnwidth]{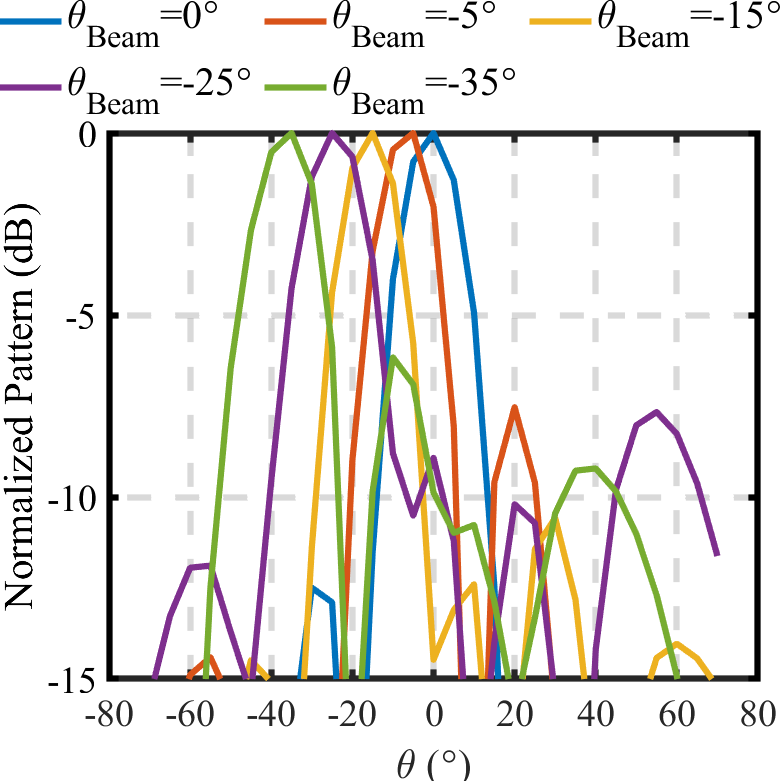}}
\par\end{centering}
\begin{raggedright}
\hspace*{0.24\columnwidth}(a)\hspace*{0.44\columnwidth} (b)
\par\end{raggedright}
\caption{Simulated scattered patterns for 3.5 GHz when the incident angle $(\theta,\phi)_{\mathrm{incident}}=(0^{\circ},0^{\circ})$,
and the main beam is steered to (a) $(\theta,\phi)_{\mathrm{beam}}=(0^{\circ},0^{\circ})$,
$(5^{\circ},0^{\circ})$, $(15^{\circ},0^{\circ})$, $(25^{\circ},0^{\circ})$,
$(35^{\circ},0^{\circ})$, (b) $(\theta,\phi)_{\mathrm{beam}}=(0^{\circ},0^{\circ})$,
$(-5^{\circ},0^{\circ})$, $(-15^{\circ},0^{\circ})$, $(-25^{\circ},0^{\circ})$,
$(-35^{\circ},0^{\circ})$.}
\label{imulated scattered pattern for 3.5 GHz}
\end{figure}

Fig. \ref{Sub-6 GHz element geometry} demonstrates the optimized
structure of the sub-6 GHz element with selected connections and fixed
RF switches. Its phase entropy for different frequencies when excited
by a vertically incident wave is shown in Fig. \ref{Performance of sub-6 GHz element}
(a), where we can find that the phase entropy can maintain a high
value, above 2.2, from $-40^{\circ}$ to $40^{\circ}$ across 3.4
to 3.55 GHz. This means that the sub-6 GHz element can provide good
reconfigurability for a wide angle and about 150 MHz bandwidth. The
specific reflection phases of the 8 reconfigurable states also prove
this, as shown in Fig. \ref{Performance of sub-6 GHz element} (b).
Thus, we obtain an sub-6 GHz element with good performance by entirely
reusing the structure of the mmWave element and without adding any
more aperture.

\section{DBI-RIS Element Integration}

After designing and prorotyping the 1-bit mmWave element and 3-bit
sub-6 GHz element, they need to be integrated together into a complete
DBI-RIS. As shown in Fig. \ref{3_5GHz_4x4_picture}, we fabricated
the sub-6 GHz RIS with a $4\times4$ element configuration. Each sub-6
GHz element was fabricated separately and mounted on an FR-4 board.
All DC feeding points were connected to the FR-4 board using metal
screws, and a DC control network was also designed on the board. Each
element featured one ground line and three control lines. In total,
48 control lines from the 16 elements were connected to the I/O pins
(3.3V output) of the FPGA, allowing for the configuration of the sub-6
GHz element states to control the scattered pattern.
\begin{figure}[t]
\begin{centering}
\textsf{\includegraphics[width=1\columnwidth]{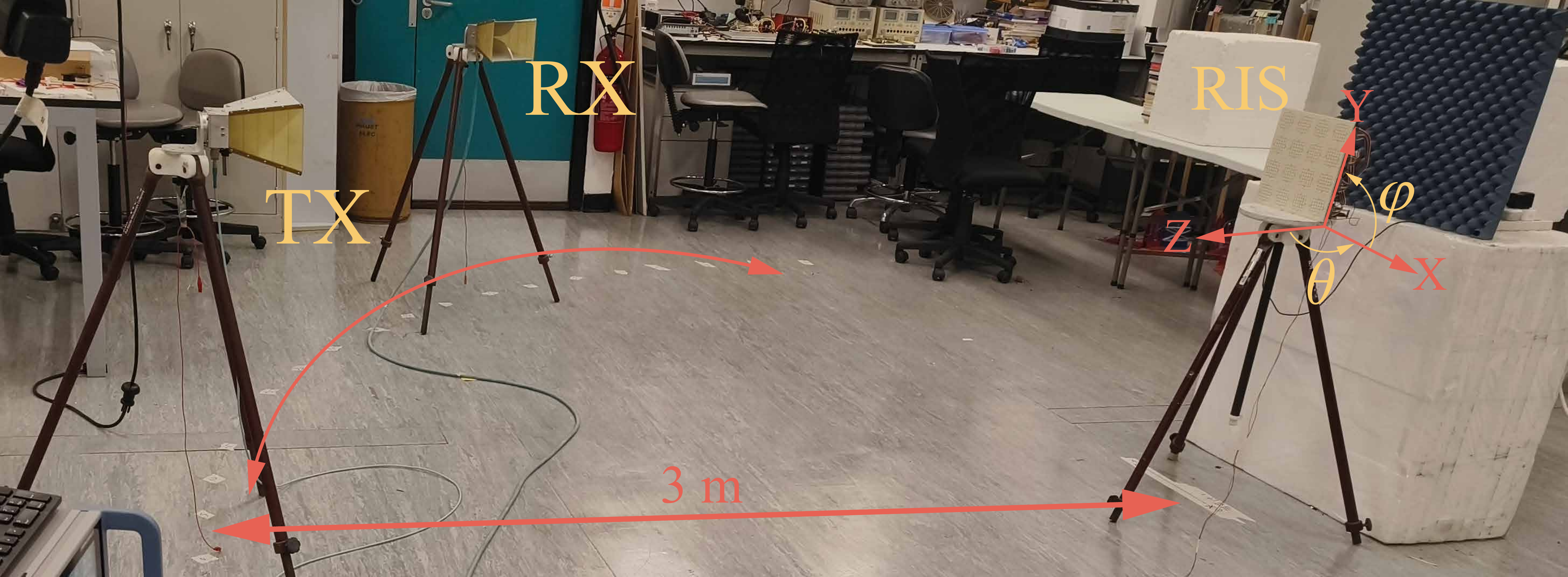}}
\par\end{centering}
\begin{centering}
(a)
\par\end{centering}
\begin{centering}
\textsf{\includegraphics[height=0.5\columnwidth]{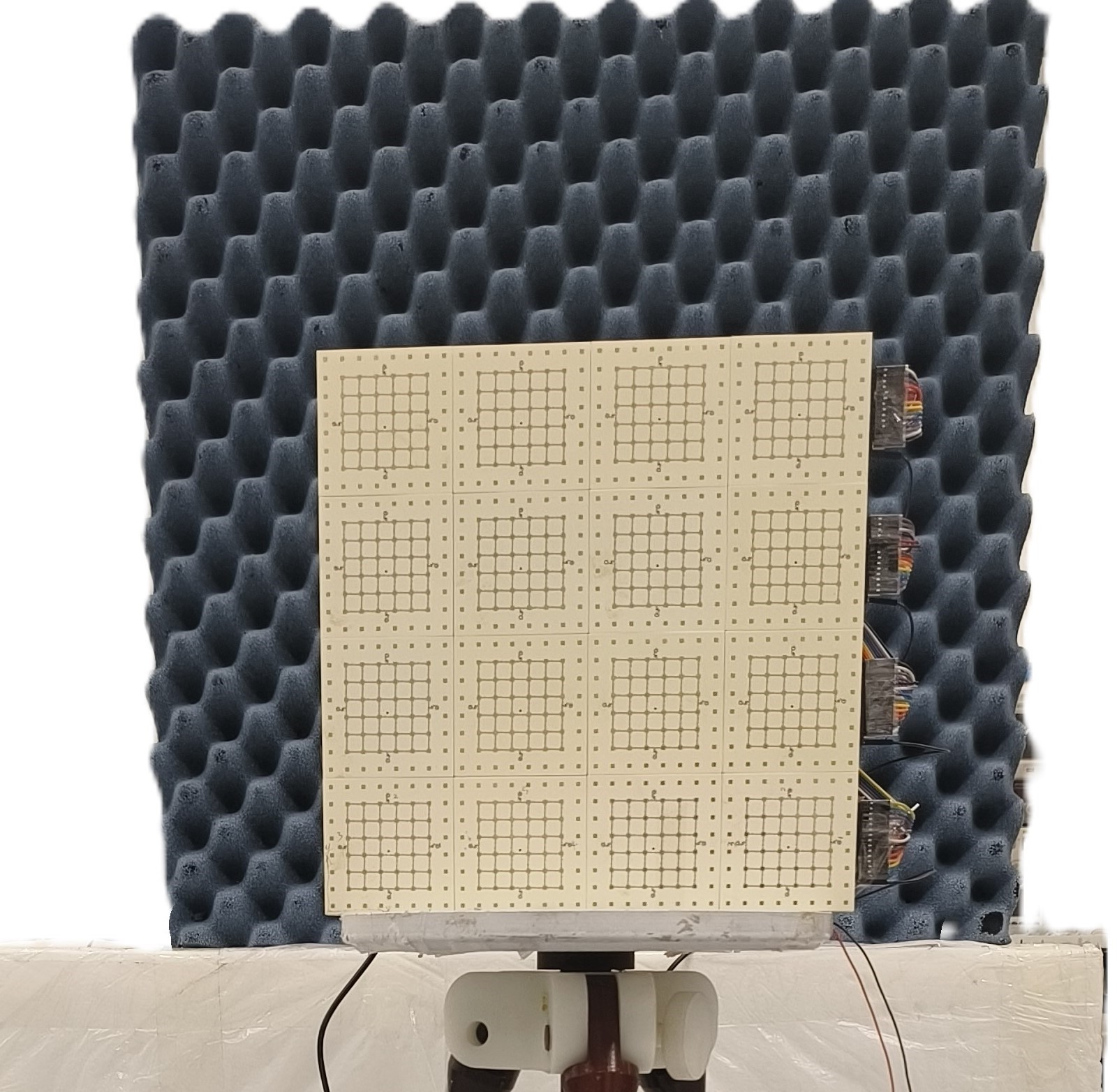}\includegraphics[height=0.5\columnwidth]{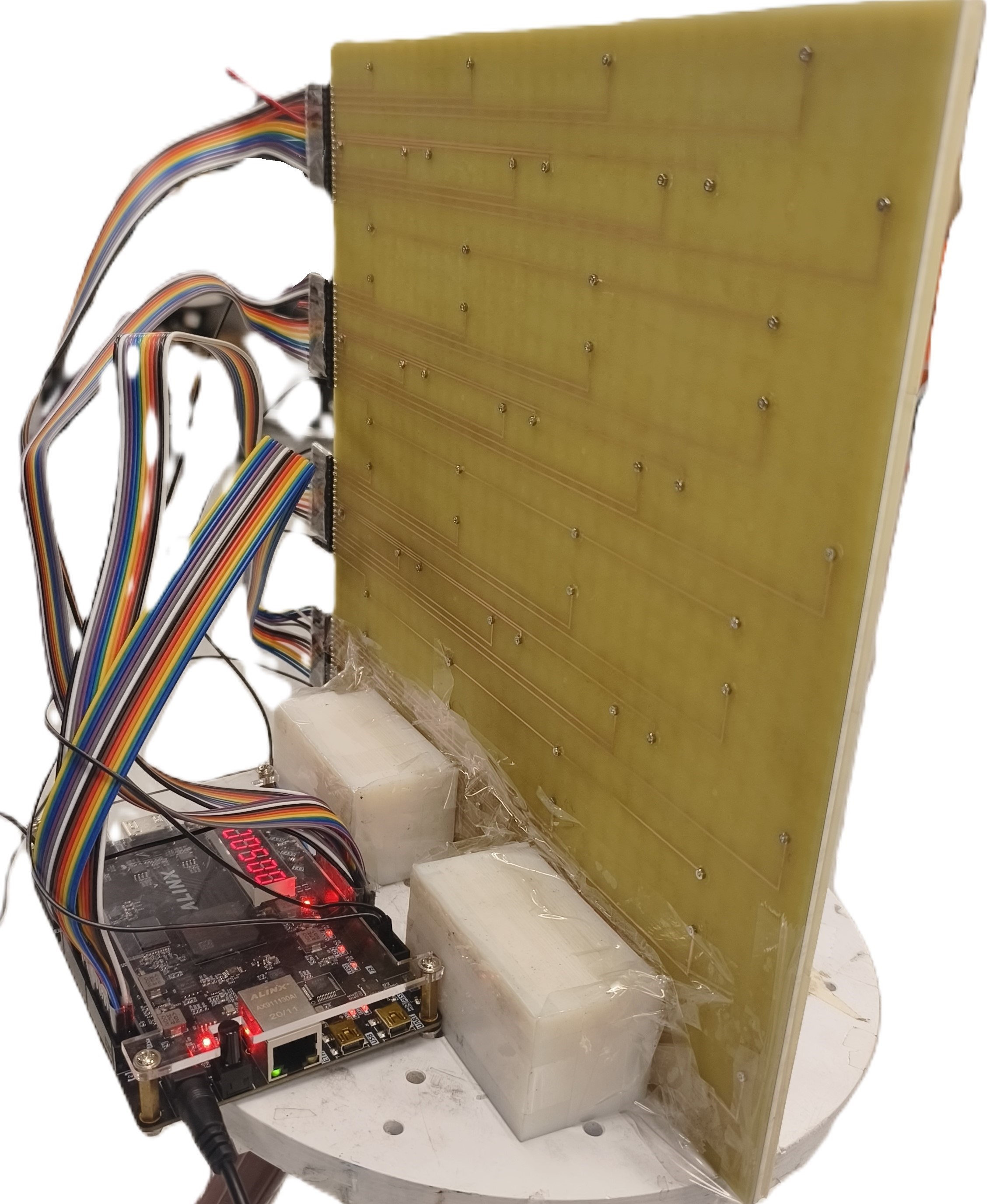}}
\par\end{centering}
\begin{raggedright}
\hspace*{0.24\columnwidth}(b)\hspace*{0.44\columnwidth} (c)
\par\end{raggedright}
\caption{(a) Photograph of the experiment setup for measuring the sub-6 GHz
scattered wave pattern of the proposed DBI-RIS. The configuration
of the coordinate system is also shown. (b) Photograph of the DBI-RIS
front, (c) Photograph of the DBI-RIS back with an FPGA controlling
system.}
\label{Photograph of the experiment setup for measuring the sub-6 GHz scatte}
\end{figure}
\begin{figure}[t]
\begin{centering}
\textsf{\includegraphics[width=0.5\columnwidth]{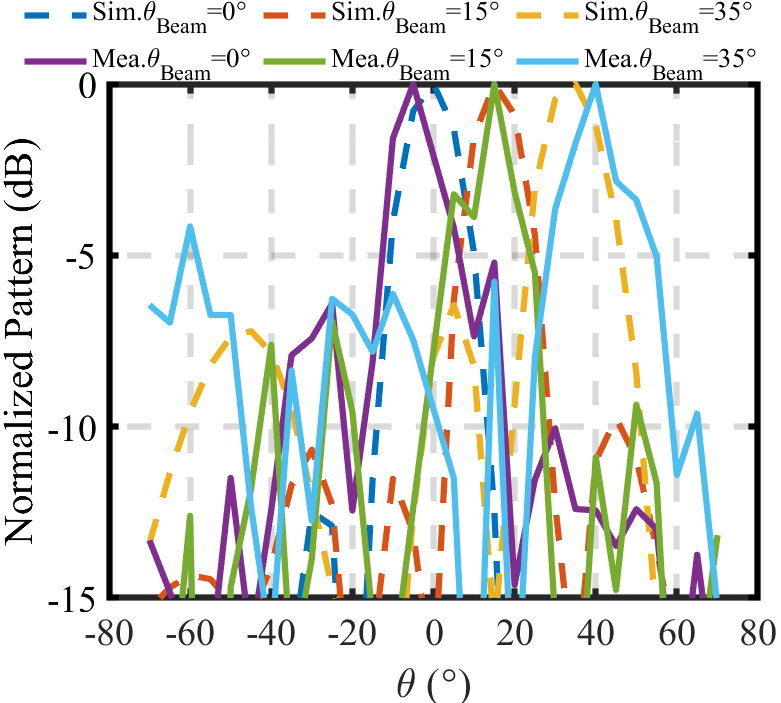}\includegraphics[width=0.5\columnwidth]{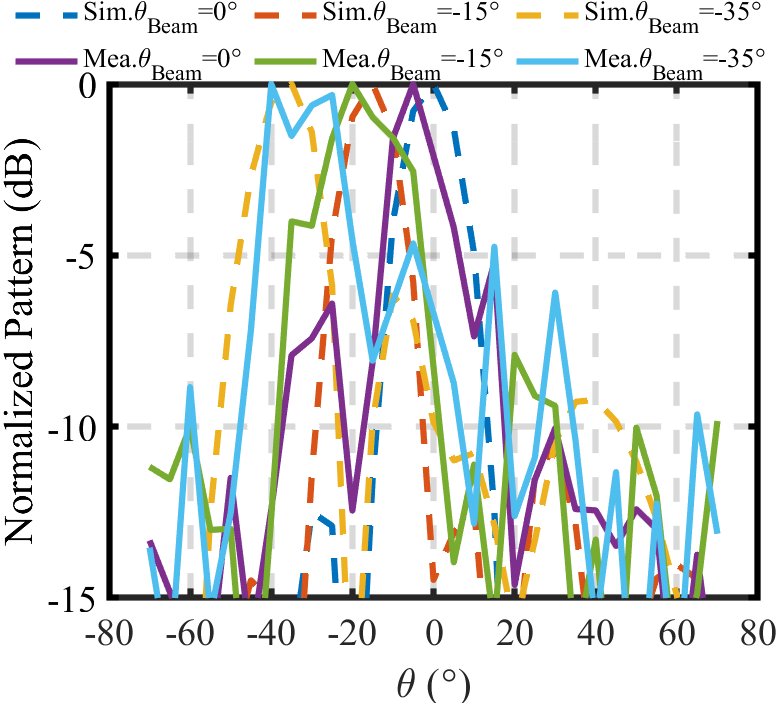}}
\par\end{centering}
\begin{raggedright}
\hspace*{0.24\columnwidth}(a)\hspace*{0.44\columnwidth} (b)
\par\end{raggedright}
\caption{Measured and simulated scattered patterns when the incident angle
$(\theta,\phi)_{\mathrm{incident}}=(0^{\circ},0^{\circ})$, and the
main beam is steered to (a) $(\theta,\phi)_{\mathrm{beam}}=(0^{\circ},0^{\circ})$,
$(15^{\circ},0^{\circ})$, $(35^{\circ},0^{\circ})$, (b) $(\theta,\phi)_{\mathrm{beam}}=(0^{\circ},0^{\circ})$,
$(-15^{\circ},0^{\circ})$, $(-35^{\circ},0^{\circ})$.}
\label{Measured and simulated scattered patterns sub-6 GHz}
\end{figure}
\begin{figure}[t]
\begin{centering}
\textsf{\includegraphics[width=0.5\columnwidth]{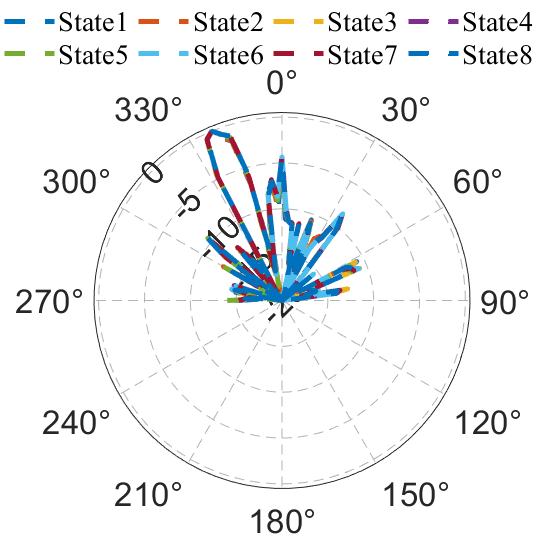}\includegraphics[width=0.5\columnwidth]{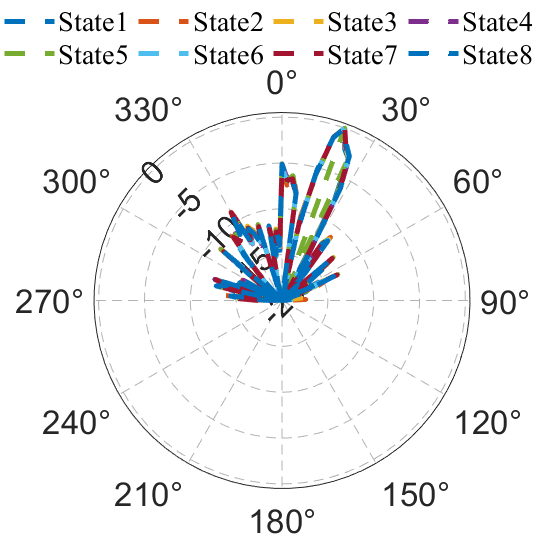}}
\par\end{centering}
\begin{raggedright}
\hspace*{0.24\columnwidth}(a)\hspace*{0.44\columnwidth} (b)
\par\end{raggedright}
\caption{The measured scattered pattern for 28 GHz and different states of
sub-6 GHz element and the main beam is steered to (a) $(\theta,\phi)_{\mathrm{beam}}=(-20^{\circ},0^{\circ})$,
(b) $(\theta,\phi)_{\mathrm{beam}}=(20^{\circ},0^{\circ})$.}
\label{mea scattered pattern of 8x8 mmwave RIS different states sub6 ghz}
\end{figure}
\begin{table*}[t]
\caption{Comparison with Related Work}
\label{table 1}
\centering{}%
\begin{tabular}{>{\centering}m{1.5cm}>{\centering}m{1.5cm}>{\centering}m{1.5cm}>{\centering}m{1.5cm}>{\centering}m{2cm}>{\centering}m{2cm}>{\centering}m{1.5cm}>{\centering}m{1.5cm}>{\centering}m{1.5cm}}
\hline 
\noalign{\vskip\doublerulesep}
Ref. & Frequency

(GHz) & Bandwidth & Num. of elements & Total size & Element reconfigurability & Num. of switches/$\lambda^{\mathbf{2}}$ & Scanning range & Num. of operation bands\tabularnewline[\doublerulesep]
\hline 
\noalign{\vskip\doublerulesep}
\hline 
\noalign{\vskip\doublerulesep}
\cite{Pan2021} & 9.3 & 2.58\% & $160\times64$ & $80\lambda\times32\lambda$ & 2 tunable states & 4 & $\pm60^{\circ}$ & Single\tabularnewline[\doublerulesep]
\noalign{\vskip\doublerulesep}
\noalign{\vskip\doublerulesep}
\cite{Zhang2020} & 5.5 & 45.4\% & $20\times20$ & $6.6\lambda\times6.6\lambda$ & 4 tunable states & 55.09 & $\pm30^{\circ}$ & Single\tabularnewline[\doublerulesep]
\noalign{\vskip\doublerulesep}
\noalign{\vskip\doublerulesep}
\cite{Wang2024} & 26.6 & 29.3\% & $20\times20$ & $7.1\lambda\times7.1\lambda$ & 2 tunable states & 8.59 & $\pm50^{\circ}$ & Single\tabularnewline[\doublerulesep]
\noalign{\vskip\doublerulesep}
\noalign{\vskip\doublerulesep}
\cite{Rao2022} & 2.4 & 4.1\% & $4\times4$ & $2.56\lambda\times2.56\lambda$ & 16 tunable states & 9.83 & $\pm50^{\circ}$ & Single\tabularnewline[\doublerulesep]
\noalign{\vskip\doublerulesep}
\noalign{\vskip\doublerulesep}
This Work & 3.5 / 28 & 4.3\% / 5\% & $4\times4$ /

$32\times32$ & $3.17\lambda\times3.17\lambda$ /

$6.35\lambda\times6.35\lambda$ & 8 tunable states /

2 tunable states & 4.8 / 1.58 & $\pm35^{\circ}$ / $\pm30^{\circ}$ & Dual\tabularnewline[\doublerulesep]
\hline 
\noalign{\vskip\doublerulesep}
\end{tabular}
\end{table*}

The scattered pattern produced by the RIS at sub-6 GHz can be calculated
analytically using the method \eqref{scattered pattern}, which is
the same approach employed in \cite{Rao2023}. Fig. \ref{imulated scattered pattern for 3.5 GHz}
displays the simulated scattered patterns for representative main
beam directions, demonstrating that the $4\times4$ sub-6 GHz RIS
is capable of steering the reflected beam within a range of -35� to
35�.

A testbed featuring two horn antennas was established to measure the
scattered patterns from the fabricated sub-6 GHz RIS, as depicted
in Fig. \ref{Photograph of the experiment setup for measuring the sub-6 GHz scatte}.
The antennas were positioned at a fixed distance of 3 m from the RIS
within the far-field region. Utilizing a vector network analyzer (Rohde
\& Schwarz ZVA40), the $S_{21}$ parameter between the transmitter
and receiver was determined. The scattered pattern was measured by
rotating the receiver in $5^{\circ}$ increments while keeping the
transmitter stationary. To account for background scattering, two
$S_{21}$ parameters were measured at each receiver position: $S_{21,env}(\Omega)$
without the RIS and $S_{21,total}(\Omega)$ with both the RIS and
environment present. The scattered RIS pattern, $S_{21,scat}(\Omega)$,
is calculated as:

\begin{equation}
S_{21,scat}(\Omega)=S_{21,total}(\Omega)-S_{21,env}(\Omega).\label{eq:9}
\end{equation}
Fig. \ref{Measured and simulated scattered patterns sub-6 GHz} compares
the calculated and measured scattered patterns in the $xoz$ plane,
focusing on six representative main beam directions while the transmitter
remains fixed along the $z$-axis. The measured pattern exhibits general
agreement with the predicted results obtained from the method \eqref{scattered pattern}.
Minor discrepancies can be attributed to factors such as fabrication
errors, non-ideal wave excitation, background scattering, and positioning
errors. The fabricated sub-6 GHz RIS successfully achieves beam steering
within the range of -35$^{\circ}$ to 35$^{\circ}$ in the $xoz$
plane. By incorporating a greater number of elements in the sub-6
GHz RIS, a wider beam steering range can be realized.

On the other hand, the fabricated 4$\times$4 sub-6 GHz RIS shares
the same aperture with 32$\times$32 mmWave RIS that can be controlled
independently. For each 8$\times$8 mmWave element, a controlling
network is designed as shown in Fig. \ref{picture 8x8 mmwave ris}
(b) and a total of 16 mmWave DC networks are combined together to
configure the function in mmWave. Since the scattered pattern of mmWave
band was already verified and measured as in section III, for simplicity,
we provided the scattered pattern by the fabricated DBI-RIS in sub-6
GHz.

To verify the independence between mmWave and sub-6 GHz elements,
we also measured the scattered pattern of the mmWave RIS under all
the 8 different states of the sub-6 GHz element, as shown in Fig.
\ref{mea scattered pattern of 8x8 mmwave RIS different states sub6 ghz}.
For any state of the sub-6 GHz element, the scattered pattern remains
almost unchanged. In other words, the state of mmWave elements and
the state of the sub-6 GHz element are totally independent to each
other, as desired.

\section{Discussion}

Table \ref{table 1} compares the proposed DBI-RIS with other related
RIS works in terms of operating frequency, bandwidth, total size,
element reconfigurability, switches density, scanning range, and the
number of operation bands. As far as we are aware, there are no other
dual-band RIS so the comparisons are with single band RIS. The authors
of \cite{Pan2021} proposed a large-scale reconfigurable reflectarray
antenna (RRA) for high gain and wide scanning angle, utilizing a fixed
feeding antenna. In contrast, \cite{Zhang2020} introduced a multifunctional
surface design capable of operating in transmission and reflection
modes, with each element offering four reconfigurable properties by
configuring the states of its six diodes. In \cite{Wang2024}, a combination
of four circular cutouts and long vias were used to achieve wideband
performance at the mmWave frequency band. Our previous work, \cite{Rao2022},
on the other hand, achieved 16 reconfigurable states using four diodes
and can enable wide scanning through a relatively small aperture.

On the other hand, all the reported related works can only operate
in a single band. Based on the shared-aperture structure-reusing method,
the proposed DBI-RIS is capable of working simultaneously and independently
at both sub-6 GHz and mmWave bands. Moreover, since the main cost
source of RIS, particularly for mmWave band RIS (more than 50\%),
comes from diodes, switch density is also an important metric for
practical implementation. The proposed DBI-RIS element can realize
more tunable states by using relatively fewer switches for both bands.
The attributes of dual-band operation and low switch density enable
the proposed DBI-RIS to bridge the research gap between existing single-band
RIS and wireless systems employing multiple bands.

\section{Conclusions}

As a conclusion, we present a novel DBI-RIS design combining mmWave
and sub-6 GHz functionalities within a single aperture to address
the research gap between single-band RISs and future multi-band wireless
systems. The mmWave element uses a double-layer patch antenna with
a 1-bit phase shifter, while the sub-6 GHz element is realized through
selectively interconnected 8$\times$8 mmWave arrays. A suspended
EBG structure and a PSI are proposed to optimize the design. Prototypes
are fabricated and experimentally verified, demonstrating successful
beam steering for both 4$\times$4 sub-6 GHz and 8$\times$8 mmWave
elements within the desired ranges, aligning well with simulated results.

\appendices{}
\begin{lyxlist}{00.00.0000}
\item [{}]~
\end{lyxlist}
\bibliographystyle{IEEEtran}
\bibliography{ref}

\end{document}